\documentclass[useAMS,usegraphicx,usenatbib]{mn2e}

\usepackage{times}

\title[On the evolutionary stage of CD-42$\bmath{\degr}$11721]{On the evolutionary stage of the unclassified B[e] star CD-42$\bmath{\degr}$11721\thanks{Based on observations done with the 1.52-m telescope at the European Southern Observatory (La Silla, Chile).}}
\author[Borges Fernandes, Kraus, Lorenz Martins \& de Ara\'ujo]{M. Borges 
Fernandes$^{1}$\thanks{E-mail: borges@ov.ufrj.br (MBF); kraus@sunstel.asu.cas.cz (MK); araujo@on.br (FXA); slorenz@ov.ufrj.br (SLM)}, M. Kraus$^{2}$, S. Lorenz Martins$^{1}$ and F. X. de Ara\'ujo$^{3}$\\
$^{1}$Observat\'orio do Valongo - UFRJ, Ladeira do Pedro Ant\^onio 43, 20080-090, Sa\'ude, Rio de Janeiro, Brazil\\
$^{2}$Astronomick\'y \'ustav, Akademie v\v{e}d \v{C}esk\'e republiky, Fri\v{c}ova 298, 251~65 Ond\v{r}ejov, Czech Republic\\ 
$^{3}$Observat\'orio Nacional-MCT, Rua General Jos\'e Cristino 77, 20921-400, S\~ao Crist\'ov\~ao, Rio de Janeiro, Brazil}
\begin{document}

\date{Accepted ********. Received ********.}

\pagerange{\pageref{firstpage}--\pageref{lastpage}} \pubyear{2006}

\maketitle

\label{firstpage}

\begin{abstract}

The star CD-42$\bmath{\degr}$11721 is a curious B[e] star sometimes 
pointed as an evolved B[e] supergiant and sometimes as a young HAeBe star, due to very uncertain or even unknown stellar
parameters, especially the distance. In this paper, we present new data gained from 
high-resolution optical spectroscopy and a detailed description of IR data of this star. We present a qualitative study of the numerous emission lines in our optical spectra and the 
classification of their line profiles, which indicate a non-spherically 
symmetric circumstellar environment. The first real detection of numerous [Fe{\sc ii}] emission lines and of many other 
permitted and forbidden emission lines is reported. From our optical data, we derive an effective temperature of $T_{\rm eff} = 14\,000\pm 1\,000$\,K, a radius of 
$R_{*} = 17.3\pm 0.6$\,R$_{\odot}$, as well as a luminosity of $L_{*} = (1.0\pm 
0.3)\times 10^{4}$\,L$_{\odot}$. We advocate that CD-42$\bmath{\degr}$11721 might be a post-main sequence object, even though a pre-main sequence nature cannot  be ruled out due to the  uncertain distance. We further found that the SED in the optical and infrared can best be fitted with an outflowing disk-forming wind scenario rather than with a spherical symmetric envelope or with a flared disc, supporting our tentative classification as a B[e] supergiant.

\end{abstract}


\section[]{Introduction}
CD-42$\bmath{\degr}$11721 (V921 Sco, Hen 3-1300, IRAS 16555-4237) is a very 
interesting galactic southern object showing the B[e] phenomenon. B[e] stars
are stars of spectral type B with an optical spectrum showing strong Balmer 
line emission in combination with a large amount of permitted Fe\,{\sc ii}
lines and forbidden O\,{\sc i} and Fe\,{\sc ii} lines. In addition, these
stars exhibit a strong near- and mid-IR excess due to circumstellar dust. 

This star was observed by Merrill \& Burwell (1949), who 
noted a peculiar nature in its spectrum. Due to the presence of a nebulosity 
around it \citep{Glass} and possible spectral and photometric variations 
\citep{Herbst, Carlson}, it was tentatively classified 
as an HAeBe star (de Winter \& Th\'e 1990; Th\'e, de Winter \& Perez 1994). Based on a possible
high temperature and luminosity, \citet{Voors} has suggested that 
CD-42$\bmath{\degr}$11721 might be a young star but not a pre-main sequence 
object, while spectral similarities among CD-42$\bmath{\degr}$11721 and 
CPD-52$\bmath{\degr}$9243 \citep[a supergiant candidate,][]{Carlson} and 
HD87643 (an unclassified B[e] star, \citealt{Hutsemekers}) increased the 
confusion about the evolutionary state of this star. An evolved nature 
was reinforced by \citet{Hutsemekers}, who claimed the existence of a nebular 
overabundance of N/O. Other indications for a post-main sequence evolutionary
phase came from the proposed high luminosity of this object by \citet{McGregor} 
and \citet{Shore}. This zoo of proposed evolutionary phases for 
CD-42$\bmath{\degr}$11721 has lead to the inclusion of this star into the list 
of unclassified B[e] stars by \citet{Lamers}. Recently, \citet{Habart} and 
\citet{Hamaguchi} suggested that CD-42$\bmath{\degr}$11721 might be member
of a cluster of several low and intermediate mass sources.

\begin{figure*}
\includegraphics[width=180mm]{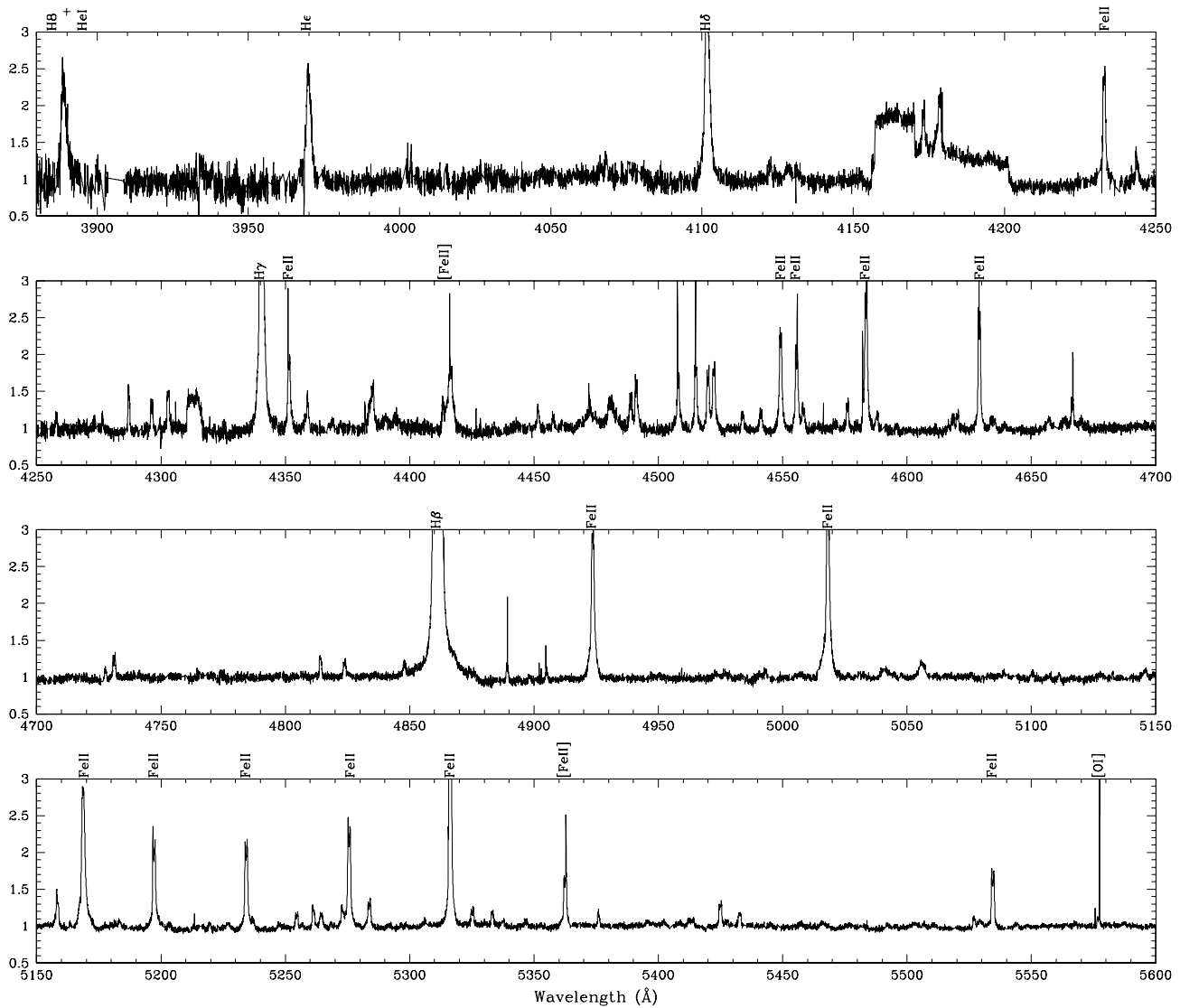}
 \vspace{0cm}
 \caption{The FEROS spectrum of CD-42$\bmath{\degr}$11721. The most intense
emission lines are indicated. A list of all identified lines is given in
Table\,\ref{Table1} in the Appendix.}
 \label{Fig.1}
\end{figure*}
                                                                                
\begin{figure*}
\includegraphics[width=180mm]{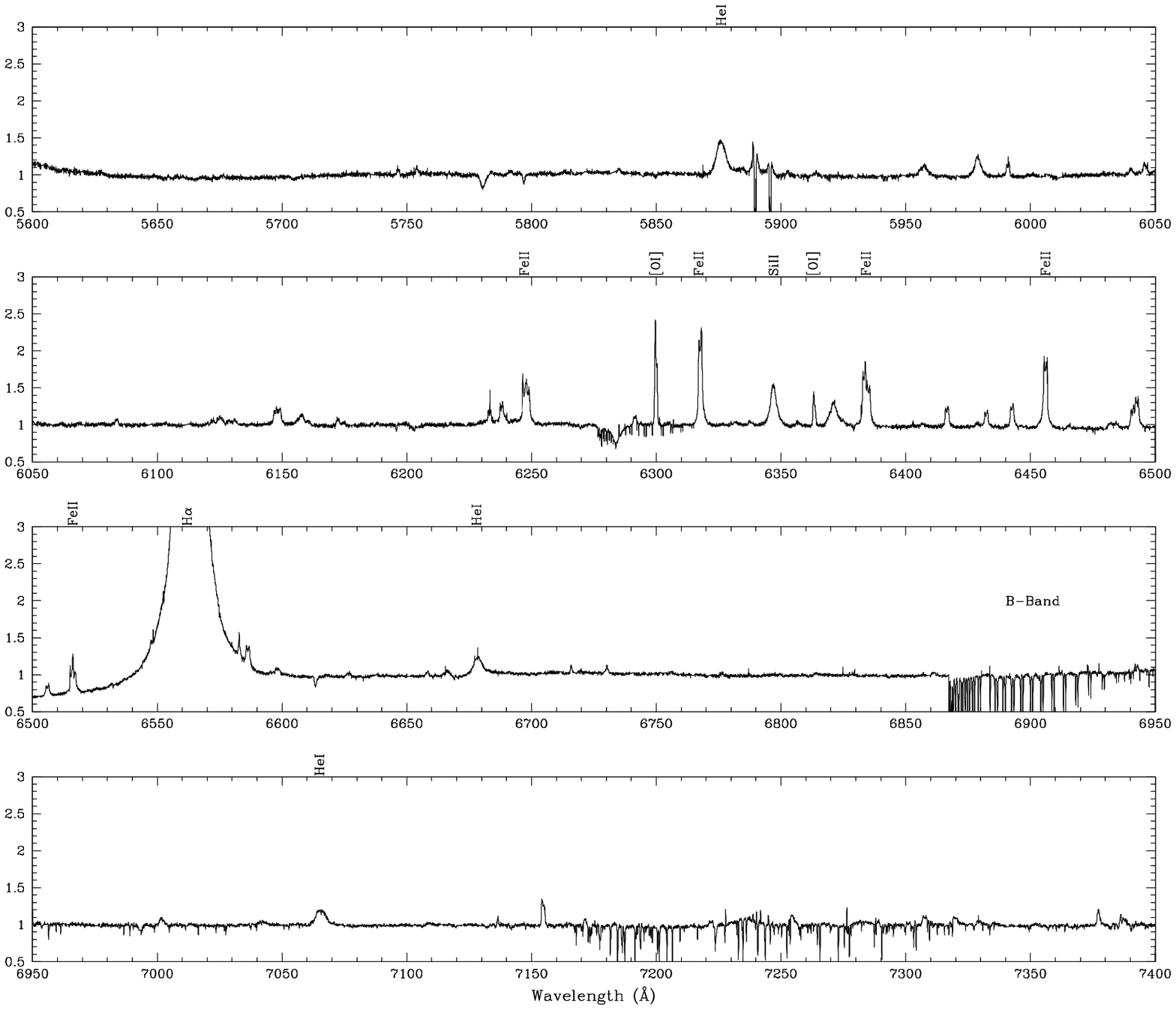}
 \vspace{0cm}
 \contcaption{}
\end{figure*}
                                                                                
\begin{figure*}
\includegraphics[width=180mm]{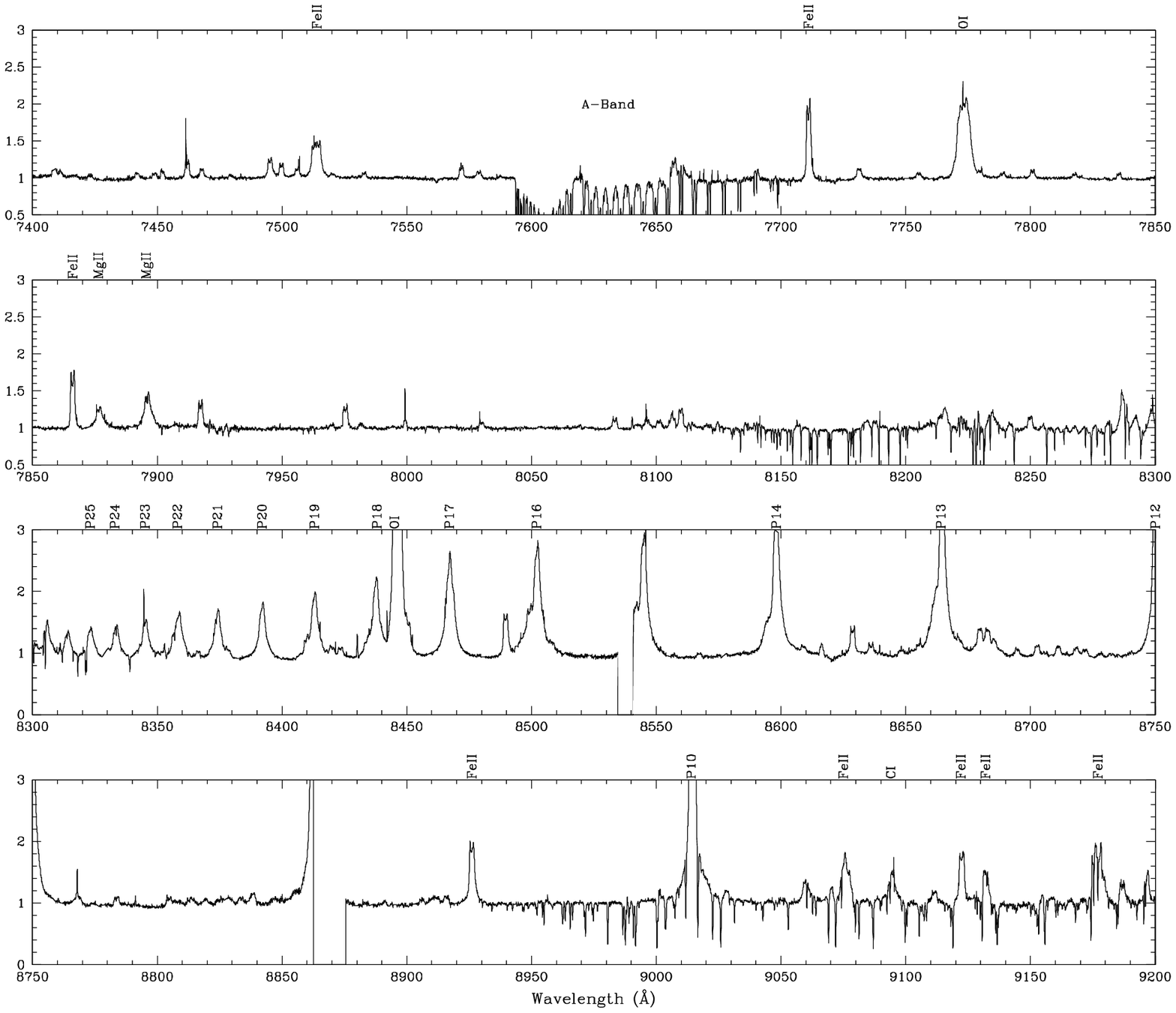}
 \vspace{0cm}
 \contcaption{}
\end{figure*}

The confusion concerning the evolutionary stage of CD-42$\bmath{\degr}$11721
is strictly linked to the absence of reliable physical parameters. Since
the optical spectrum does not display any photospheric lines, the spectral type 
derived  using different sets of observations and methods,
varies between B0 and Aep,
and consequently the T$_{\rm eff}$ ranges from 31\,600\,K to 12\,300\,K 
\citep{Hillenbrand, Cidale}. The range in distance
is even worse: we found values from 136 pc up to 2.6 kpc \citep{Shore, deWinter,
Elia}. Consequently its luminosity is not known either, being in the range 
$1.9 < \log(L/L_{\sun}) < 4.9$ (McGregor et al. 1988; Shore et al. 1990, Hillenbrand et al. 1992).
These uncertain stellar parameters and therefore the rather poor knowledge of 
the evolutionary phase of CD-42$\bmath{\degr}$11721 have stimulated us to 
investigate this object in more detail, based on a multiwavelength analysis.

In this paper, we are presenting new high- and low-resolution optical spectra together with a more detailed description of available infrared data taken from both the public catalogues as well as from the literature. We obtained a 
detailed optical spectral atlas, where the zoo of emission lines (many of them previously undetected) and the variety in line profiles give hints for the gas dynamics and distribution in the close-by circumstellar medium around CD-42$\bmath{\degr}$11721. We could also derive a set of stellar parameters based on our high-resolution data, allowing to compare the position of CD-42$\bmath{\degr}$11721 in the HR diagram with evolutionary tracks. Considering these stellar parameters, we have also modelled its SED considering different circumstellar scenarios: a spherically symmetric geometry, a passive flared disc and also an outflowing disc wind.
   
The structure of the paper is as follows. In Sect.\,\ref{obs} we describe our observations, the data reduction and the public data used. Sect.\,\ref{atlas} gives a detailed optical spectral atlas and a description of the zoo of emission lines and the variety in line
profiles found. In Sect. 4 we present a description of the IR 
region. On the basis of our optical spectra we derive in Sect.\,\ref{parameters} the stellar parameters. In Sect. 6 we describe the SED 
modelling of this star, presenting the numerical codes developped by us, the 
results obtained and also the comparison between them and the literature ones. In Sect. 7, we discuss the possible nature of CD-42$\bmath{\degr}$11721 and present the conclusions of our study.

\section[]{Observations \& reductions}\label{obs}

Our optical spectra were obtained at the ESO 1.52-m telescope in La Silla (Chile) using 
the high resolution Fiber-fed Extended Range Optical Spectrograph (FEROS) and 
the low resolution Boller \& Chivens spectrograph.  

FEROS is a bench-mounted Echelle spectrograph with fibers, that cover a sky 
area of 2$\arcsec$ of diameter, with a wavelength coverage 
from 3600\,\AA \ to 9200\,\AA \ and a mean spectral resolution of R = 48\,000 
corresponding to 2.2 pixels of 15\,$\mu$m. It has a complete automatic 
on-line reduction, which we adopted. Our FEROS spectrum presents the highest 
resolution used for observations of CD-42$\bmath{\degr}$11721 to date. It was 
taken on June 10, 2000, with an exposure time of 5400 seconds. The S/N ratio in 
the 5500\,\AA \ region is approximately 60. This high resolution spectrum and long wavelength coverage, shown in Fig.\,\ref{Fig.1} will  be used  
to obtain a detailed optical spectral atlas, for a detailed description of the different line profiles, to derive equivalent widths, and also to deblend some features in the low resolution spectra. 

The low resolution Boller \& Chivens spectrum (hereafter Cassegrain spectrum) 
was taken on the same observation run at ESO, but on June 9, 2000, with an 
exposure time of 900 seconds. The instrumental setup employed made use of 
grating \#23 with 600 l mm$^{-1}$, providing a resolution of $\sim 4.6$\,\AA \ 
in the range 3800-8700\,\AA. In the 5500\,\AA \ continuum region, the S/N ratio 
of the Cassegrain spectrum is aproximately 180. Since there is no completely 
line free region in the spectrum, the S/N derived is an upper limit. The slit 
width used has 2$\arcsec$. Considering that the nebula in which 
CD-42$\bmath{\degr}$11721 is embedded has dimensions of $\sim$ 40$\arcsec$ 
$\times$ 80$\arcsec$ \citep{Hutsemekers}, our observations were
centered on the innermost parts of it. The Cassegrain spectrum was reduced 
using standard IRAF tasks, such as bias subtraction, flat-field normalization, 
and wavelength calibration. We have done absolute flux calibration using 
spectrophotometric standards from \citet{Hamuy}. This spectrum is used 
to derive intensities of the emission lines for which we 
perform a detailed quantitative analysis.

Equivalent widths and line intensities in the linearized spectra have been 
measured using the IRAF task that computes the line area above the adopted 
continuum. Uncertainties in our 
measurements come mainly from the position of the underlying continuum and we 
estimate the errors to be about 20\,\% for the weakest lines (equivalent widths lower than 2\AA) and about 10\,\% for the strongest lines. 

We have also obtained photometric data from the literature, intending to describe the complete SED of CD-42$\bmath{\degr}$11721. In addition, we have described the IR spectra obtained by ISO and IRAS satellites. These spectra are public and can be downloaded in the following web addresses: SWS01-ISO - http://isc.astro.cornell.edu/$\sim$sloan/library/swsatlas/aot1.html; LRS-IRAS - 
http://www.iras.ucalgary.ca/$\sim$volk/getlrs$\_$plot.html. The IRAS LRS spectrum covers a small wavelength range, from 7.5 to 22\,$\mu$m, with a spectral resolution of about 20-60. On the other hand, the ISO SWS01 spectrum has a higher resolution of $\sim$ 1500 and a wavelength coverage from 2 to 50\,$\mu$m.

\section[]{The optical spectral atlas: emission lines and their profiles}\label{atlas}

To identify the lines and to create a spectral atlas of the optical 
region of CD-42$\bmath{\degr}$11721, we have used the line lists provided by
\citet{Moore, Thackeray, Hamann, McKenna} 
and \citet{Landaberry}. We have also looked up two sites on
the web: NIST Atomic Spectra Database Lines Form 
({\it http://physics.nist.gov/cgi-bin/AtData/lines$\_$form}) and The Atomic 
Line List v2.04 ({\it http://www.pa.uky.edu/$\sim$peter/atomic/}).

Fig.\ref{Fig.1} shows the complete high resolution spectrum of 
CD-42$\bmath{\degr}$11721. It is completely dominated by emission lines. The 
only absorption lines visible are the Ca\,{\sc ii} and Na\,{\sc i} lines, which are of interstellar origin, as well as the Diffuse Interstellar Bands (DIBs). Our spectra do not exhibit any photospheric lines.

A huge number of permitted and forbidden emission lines were identified in the 
FEROS spectrum of CD-42$\bmath{\degr}$11721 and listed in Table\,\ref{Table1}
given in the Appendix. The individual columns in Table\,\ref{Table1}
contain  the observed wavelength (Col. 1), equivalent width (W($\lambda$), in \AA, 
Col. 2), the line profile type (Col. 3), and the proposed identification 
(Col. 4) for each line. The line identification given in Col. 4 encloses the ionization state of the 
element with the proposed transition and multiplet as well as the rest 
wavelength of the transition. It is possible that more than one ion can be 
allocated to a single feature. In these cases, we give some possible 
alternative identifications. For some lines no identification could be found. 
These lines remain unidentified, labelled as ''Uid`` in the tables.

Inspection of Table\,\ref{Table1} reveals a huge zoo of emission lines
from atoms mostly singly ionized, as well as a variety in their
line profiles. Interestingly, there are also many emission lines from neutral metals like 
C{\sc i}, N{\sc i}, and O{\sc i}, and some sporadic traces for the emission
of S{\sc i}, and Mg{\sc i} as well. Many of the lines listed in
Table\,\ref{Table1} are detected by the first time.
 
In the following, we briefly discuss the presence of individual elements and 
lines and, if possible, compare our observations with previously published data. 
We especially highlight deviations of our observations from data in the 
literature which all seem to point in favour of either central source 
variability or ongoing changes in the CSM conditions. In addition, we classify 
the different line profiles.

\subsection{The zoo of emission lines}

\begin{figure}
\includegraphics[width=85mm]{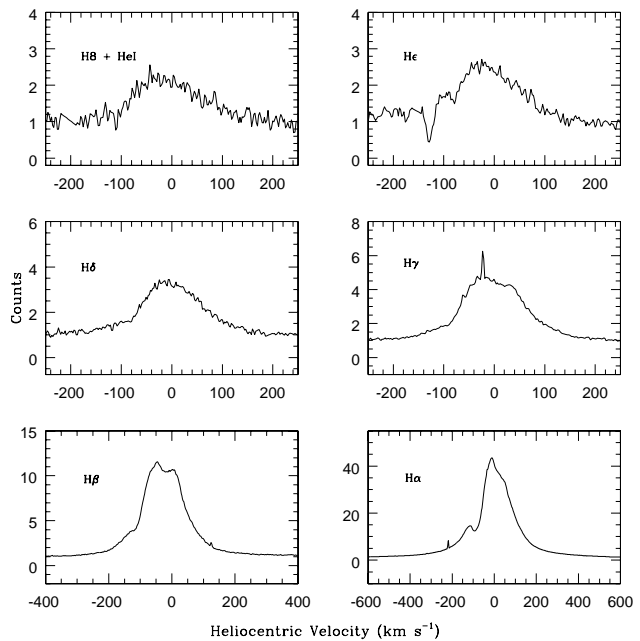}
\caption{Balmer lines from the CD-42$\bmath{\degr}$11721 high resolution 
(FEROS) spectrum. The lines are broad and in emission with H$\alpha$, H$\beta$ 
and H$\gamma$ showing indications for a double-peaked profile. The absorption 
dip within the blue wing of H$\epsilon$ is the interstellar Ca\,{\sc ii} 
3968\,\AA~line. Due to the scale of these figures, the extended wings of 
especially the H$\alpha$ and H$\beta$ lines cannot be seen.}
\label{balmer}
\end{figure}

\begin{figure}
\includegraphics[width=80mm]{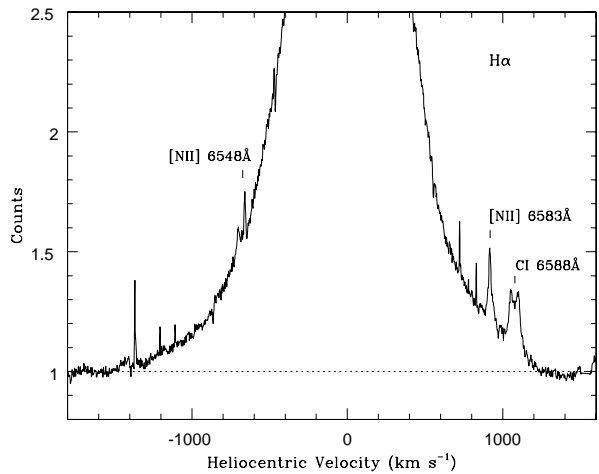}
 \caption{The wings of the H$\alpha$ line, extending to $\sim
1400$\,km\,s$^{-1}$. The C{\sc i} $\lambda$ 6588 as well as the [N\,{\sc ii}] $\lambda$$\lambda$ 6548, 6583 lines are
also indicated. The dotted line represents the continuum.}
 \label{Halpha_wings}
\end{figure}

{\bf Hydrogen:}
The strongest lines in the spectra of CD-42$\bmath{\degr}$11721 are the Balmer 
lines (Fig.\,\ref{balmer}). We could identify the Balmer series up to H9 at the 
blue edge of our spectrum. While H$\alpha$, H$\beta$ and H$\gamma$ show 
indications for a double-peaked profile and broad wings extending to 
velocities of about 1400, 600, and 200\,km\,s$^{-1}$ respectively (for the H$\alpha$ wings 
see Fig.\,\ref{Halpha_wings}), the higher Balmer lines are singly peaked and 
their wing velocities decrease systematically. The equivalent width of H$\alpha$ is almost a factor 2.5 lower than the value reported by Lopes, Damineli Neto \& de Freitas Pacheco (1992).
The H$\beta$ equivalent width has also been reported to be 
variable \citep{Lopes}. While it was found to be 27.7\,\AA~in the 1985 
observations, it increased to 41.5\,\AA~in the 1988 observations. Our value,
observed in 2000, lies with 30.4\,\AA~in between.

In addition, at the long-wavelength end of the spectrum, the Paschen series
arises. From this series we could clearly identify the lines from Pa(10) up to
Pa(40). Their profiles are all narrow and single-peaked (see Fig.\,\ref{Fig.1}).

{\bf Helium:}
Helium seems to be variable in the spectrum of CD-42$\bmath{\degr}$11721.
While it had (possibly) been detected in the years 1945--1948 by 
\citet{Merrill}, it was absent in the spectra of \citet{Carlson} taken in
1949--1951 and 1962, and reappeared again in the spectra observed in
1985 by \citet{Lopes} and 1989 by \citet{Hutsemekers}. And also our data 
which have been taken in 2000 show emission of He{\sc i}. We do not detect 
any line from He{\sc ii}.

{\bf Oxygen:}
The detection of the [O{\sc i}] 6300\,\AA~was first reported by \citet{Acke2}.
This line (and other oxygen lines) have not been seen during earlier
observations performed by \citet{deWinter} and \citet{Hutsemekers}.

\begin{figure}
\includegraphics[width=85mm]{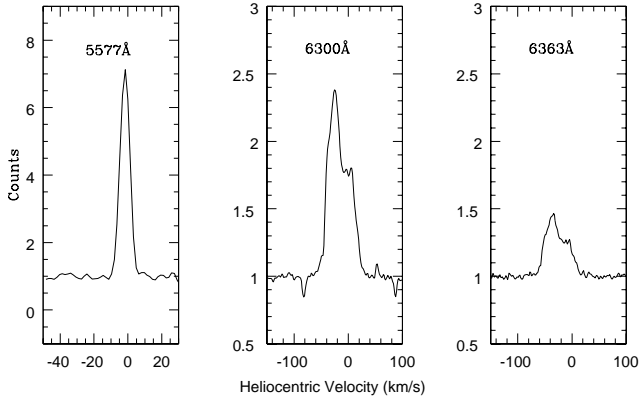}
 \caption{The [O\,{\sc i}] lines at 5577 \AA, 6300 \AA \ and 6363 \AA. Except 
the $\lambda$ 5577 line, these lines are double peaked with a missing red peak.}
 \label{ox1_lines}
\end{figure}

Our data, which have been taken two years earlier than those of \citet{Acke2}, 
confirm the presence of this emission line, and we can report on numerous 
additional identifications of emission lines of neutral oxygen, both forbidden 
and permitted. Interestingly, the [O{\sc i}] 6300\,\AA~and
6364\,\AA~lines show an asymmetric profile resembling a double-peaked line
with a missing red peak, while the 5577\,\AA~line is clearly single-peaked and
very narrow compared to the other two lines (Fig.\,\ref{ox1_lines}).
The profile of the [O{\sc i}] line $\lambda$ 6300 shown by \citet{Acke2}
agrees with ours, but there are significant differences in the equivalent width
of this line: while \citet{Acke2} derived a value of 1.77\,\AA,~the equivalent 
width found from our observations is only 1.20\,\AA.~Since the set-up for both 
observations was quite similar, this difference in equivalent width might hint 
towards rapidly changing conditions, e.g. in density and temperature structure,
within the line forming region. 

In addition, we found (though very weak) the previously not detected [O{\sc ii}] lines
$\lambda\lambda$ 7319, 7330. No single O{\sc iii} line was identified.

{\bf Nitrogen:}
The only previously detected emission line from nitrogen in the spectrum of 
CD-42$\bmath{\degr}$11721 is the line [N{\sc ii}] $\lambda$\,5755, cited by 
\citet{deWinter} and \citet{Hutsemekers}. The latter, however, ascribed the 
emission to the nebula and could not identify this line in the stellar 
spectrum. Our high-resolution spectrum confirms the existence
of this emission line. In addition, we detected the lines [N{\sc ii}] $\lambda\lambda$ 6548, 6584, 
as well as N{\sc ii} $\lambda$ 6173. Remarkably, we also identified many 
permitted emission lines from N{\sc i}.  

{\bf Carbon:}
The detection of any carbon line has not been reported yet. Our spectrum
clearly shows the presence of several lines of C{\sc i} (see e.g. 
Fig.\,\ref{Halpha_wings}), one line of C{\sc ii} and C{\sc iii} each, as well 
as the possible detection of a [C{\sc i}] line.

{\bf Sulfur:}
The existence of the [S{\sc ii}] $\lambda\lambda$ 6716, 6731 lines has formerly 
been reported by \citet{Hutsemekers} as coming from the nebula.
In addition to these lines, our spectrum also contains the line [S{\sc ii}] $\lambda$ 4068
while the $\lambda$ 4076 line is missing. From 
S{\sc iii} we could identify only one line, i.e. [S{\sc iii}] $\lambda$ 9069. 
The line [S{\sc iii}] $\lambda$ 6311 does not show up in our spectrum. However,
as in the case of N, O, and C, we have some indications for the presence of 
S{\sc i} emission.

{\bf Iron:}
It is the element with by far the largest number of emisison lines. With the
exception of four lines allocated to Fe{\sc iii} (two forbidden and two 
permitted), all lines are from singly 
ionized iron. CD-42$\bmath{\degr}$11721 is known to show a huge number of 
Fe{\sc ii} lines \citep[e.g.][]{Carlson, Hutsemekers}, however, the existence 
of an equally large amount of [Fe{\sc ii}] lines as we found in our spectrum,
has previously not been reported. Actually \citet{deWinter} reported on a possible detection of a [Fe{\sc ii}] line, however this identification is rather doubtful due to their low spectral resolution. These lines show either single-peaked, double-peaked, or even multiple-peaked profiles, 
indicating a high complexity of the CSM. The permitted lines are usually more 
intense than the forbidden ones and also broader with wings extending to 
100--200\,km\,s$^{-1}$, while the line wings of the forbidden lines indicate 
velocities of about $\sim 50-60$\,km\,s$^{-1}$, only.

\begin{table}
\caption{Velocities of the H Balmer lines and some forbidden lines seen in the 
FEROS spectrum. Listed are the ions, the laboratory wavelength of the line, the 
blue and red wing velocities, the mean expansion velocity, and the line center 
velocity.}
\begin{tabular}{cccccc}
\hline
Ion & $\lambda$ & $v_{\rm blue}$ & $v_{\rm red}$ & $<v_{\rm exp}>$ & $v_{0}$ \\
 & [\AA] &   [km s$^{-1}$]  & [km s$^{-1}$] & [km s$^{-1}$] & [km s$^{-1}$] \\
\hline
H$\alpha$     & 6562 & -1500 & +1300 & 1400 & -100 \\
H$\beta$      & 4861 & -500 & +600 & 550 & +50 \\
H$\gamma$     & 4340 & -200 & +200 & 200 & 0 \\
H$\delta$     & 4101 & -150 & +150 & 150 & 0 \\
H$\epsilon$   & 3970 & -140 & +140 & 140 & 0 \\
H8          & 3889 & -120 & +120 & 120 & 0 \\
\hline
O{\sc i}    & 6300 & -80 & +40 & 60 & -20 \\
O{\sc i}    & 6364 & -70 & +20 & 45 & -25 \\
S{\sc ii}   & 4068 & -60 & +30 & 45 & -15 \\
S{\sc ii}   & 6716 & -50 & +30 & 40 & -10 \\
S{\sc ii}   & 6731 & -80 & +30 & 55 & -25 \\
N{\sc ii}   & 5755 & -80 & +20 & 50 & -30 \\
N{\sc ii}   & 6584 & -60 & +10 & 35 & -25 \\
Cr{\sc ii} & 8106 & -90 & +40 & 65 & -25 \\
Cr{\sc ii} & 8110 & -80 & +40 & 60 & -20 \\
\hline
\end{tabular}
\label{Table_2}
\end{table}

{\bf Other elements:}
In addition, we identified emission lines from the following metal ions:
Cr{\sc ii}, Mg{\sc i}, Mg{\sc ii}, Mn{\sc ii}, Ti{\sc ii}, Si{\sc ii}, and 
Ne{\sc ii}. Most of them have previously not been detected. 

\subsection{Classification of the line profiles}

To understand the structure of the CSM of CD-42$\bmath{\degr}$11721, we sort the observed emission line profiles into 
three categories (see Table\,\ref{Table1}): (i) single-peaked, (ii) 
double-peaked, and (iii) multiple-peaked. In agreement with the descriptions in 
earlier studies \citep[e.g.][]{Lopes}, no single P Cygni profile was detected.

{\bf Single-peaked profiles:}
These profiles clearly dominate. They can be found without any exception for 
every element in every ionization stage, independent whether these are 
forbidden or permitted lines. To draw any conclusion about the 
dynamics of the CSM we measured the line wing velocities of several forbidden
(and therefore optically thin) emission lines. The results are listed in
Table\,\ref{Table_2}, together with the high wing velocities found from the 
Balmer lines. Two striking conclusions can be drawn from this table:
(i) the mean wing velocity (which we call the expansion velocity, $v_{\rm
exp}$) is on the order of 50--60\,km\,s$^{-1}$ and about constant for all
the forbidden lines, and
(ii) the line center velocity of all lines is also about constant and on the
order of -20\,km\,s$^{-1}$.

{\bf Double-peaked profiles:}
There are quite a lot of lines showing a double-peaked profile. The blue and 
red peaks are thereby of equal strength with only some rare exceptions. 
One remarkable exception are the [O{\sc i}] lines at 6300\,\AA~and 
6364\,\AA~for which the red peak is missing (Fig.\,\ref{ox1_lines}). Another 
exception are the Balmer lines, especially H$\alpha$ and H$\beta$ (see 
Fig.\,\ref{balmer})

\begin{table}
\caption{Peak separation of double-peaked lines for different ions.}
\begin{tabular}{ccccccc}
\hline
Ion &  line &   peak separ. & $\vert$  & Ion &  line &   peak separ.\\
 &  [\AA] &   [km s$^{-1}$] &  $\vert$  &   &  [\AA] &   [km s$^{-1}$]\\
\hline
H$\alpha$ & 6562 & 130 & $\vert$   &  Fe{\sc ii} & 5169 & 40 \\
H$\beta$  & 4861 & 60 & $\vert$   &  Fe{\sc ii} & 5197 & 50 \\
$ $[O{\sc i}] & 6300 & 40 & $\vert$  &  Fe{\sc ii} & 5234 & 40 \\
$ $[O{\sc i}] & 6364 & 40 & $\vert$  &  Fe{\sc ii} & 5276 & 40 \\
$ $[Fe{\sc ii}] & 5283 & 40 & $\vert$  &  Fe{\sc ii} & 5316 & 40 \\
$ $[Fe{\sc ii}] & 7155 & 40 & $\vert$  &  Fe{\sc ii} & 5534 & 40 \\
Fe{\sc ii} & 4520 & 30 & $\vert$   & Fe{\sc ii} & 6318 & 40 \\
Fe{\sc ii} & 4549 & 40 & $\vert$   & Fe{\sc ii} & 7711 & 50 \\
Fe{\sc ii} & 4923 & 30 & $\vert$   & Fe{\sc ii} & 7866 & 50 \\
Fe{\sc ii} & 5018 & 40 & $\vert$  &  &  &   \\
\hline
\end{tabular}
\label{peak_sep}
\end{table}

In Table\,\ref{peak_sep} we list some of the well resolved double-peaked
lines from the FEROS spectrum together with their peak separation, which lies 
between 30 and 50\,km\,s$^{-1}$ (neglecting the Balmer lines).

Interestingly, the double-peaked profiles are present only for neutral
metals (like O{\sc i}, N{\sc i}, C{\sc i}, and S{\sc i}) and ions with low ionization potential, i.e. lines from Fe{\sc ii}, Mg{\sc ii}, and Cr{\sc 
ii} with $\chi < 8$\,eV, which is well below the ionization potential of 
hydrogen. This fact has two severe implications:

\begin{itemize}
\item the emitting material must be neutral in hydrogen, and
\item the double-peaked profiles might indicate rotation. 
\end{itemize}

Combination of both seems to suggest the existence of a neutral (in hydrogen)
rotating disc or an equatorial outflowing wind (see e.g. Zickgraf 2003).
However, we emphasize that the double-peaked profiles are
not restricted to permitted lines only, but are also present in several
forbidden lines from [O{\sc i}], [Cr{\sc ii}], and of course [Fe{\sc ii}].

{\bf Multiple-peaked profiles:}
These types of profiles are found especially for permitted emission lines of 
Fe{\sc ii} where three peaks or 
more are observed. The existence of such multiple-peaked lines 
favours a highly complex multi-component emission region.

\subsection{Comparison with optical spectra of other B[e] stars}

\begin{figure}
\resizebox{\hsize}{!}{\includegraphics{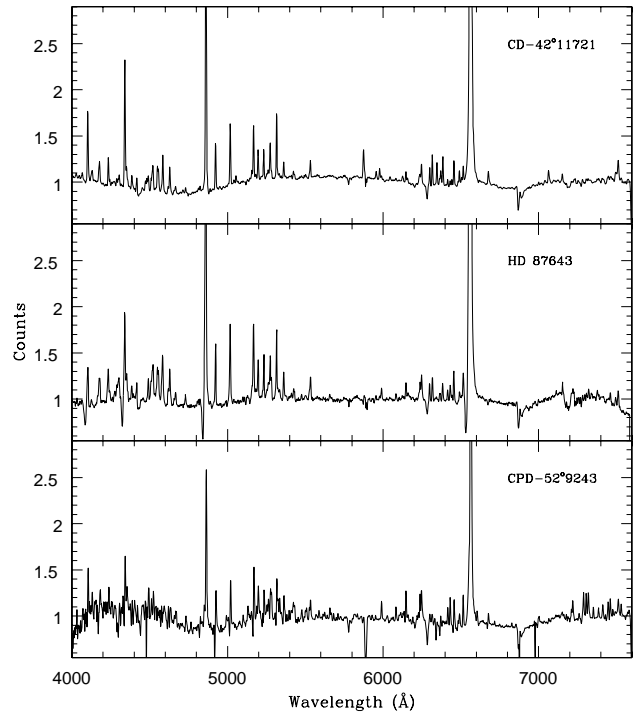}}
 \caption{Comparison of the Cassegrain spectra of CD-42$\bmath{\degr}$11721
with some other B[e] stars. Though quite similar at first glance, they show 
severe differences: the spectrum of HD\,87643 does show the presence of P Cygni
profiles, and the spectrum of CPD-52$\bmath{\degr}$9243 has the He{\sc i} lines 
in absorption.} 
 \label{compare}
\end{figure}

Based on low-resolution optical observations, it has been noted in the 
literature that the spectrum of CD-42$\bmath{\degr}$11721 is quite similar
to some other B[e] stars. One of these is HD\,87643 (Hutsem\'ekers \& van Drom 1990; Oudmaijer et al. 1998), another 
object belonging to the group of unclassified B[e] stars since it also shows
characteristics that might classify this object
either as a HAeB[e] stars or as a B[e] supergiant \citep{Lamers}.
The low-resolution spectra as shown in Fig.\,\ref{compare} are indeed 
{\sl similar}, however, there exist also differences that show up already in this 
plot: in contrast to CD-42$\bmath{\degr}$11721, the spectrum of HD\,87643
clearly shows the presence of P Cygni profiles.

Another object found to exhibit a similar optical low-resolution spectrum
(also included in Fig.\,\ref{compare}), is CPD-52$\bmath{\degr}$9243
(Carlson \& Henize 1979; Swings 1981; Winkler \& Wolf 1989), a galactic B[e] supergiant candidate \citep{Lamers}.
The most obvious difference in its optical spectrum is, however, the presence 
of the He{\sc i} lines in {\sl absorption}.

For these three stars, high-resolution (FEROS) observations show very clearly that there 
exist numerous other differences (see Borges Fernandes 2004), which cannot be listed here in detail.

\section{The infrared region: gas and dust features}

Concerning to the IR region, CD-42$\bmath{\degr}$11721 was observed by ISO and IRAS satellites. The IRAS LRS spectrum (Fig. 6) is characterized by a strong absorption centered around 
10\,$\mu$m, originated by amorphous silicate, that is not seen in the ISO SWS01 
spectra. It was classified by Zhang, Chen \& He (2004) as {\it "noisy or 
incomplete"}, meaning that the absorption could not be real. Actually, the first 
part of LRS IRAS (from 7.6 to 13.4 $\mu$m) is not so noisy and we can clearly 
see the previously known as "Unindentified IR bands" (UIR) that are commonly 
accepted nowadays as being caused by vibrational transitions of Polycyclic 
Aromatic Hydrocarbons (PAH), at 7.6, 11.3 and 12.7 $\mu$m.  

The SWS01 ISO spectrum (see also Fig. 6) of this object shows especially a strong rising continuum 
beyond 15 $\micron$, being clearly seen gas and dust emission features (Voors 
1999). Concerning the gas emission is remarkable the presence of many H 
recombination lines (Benedettini et al. 1998) and also the presence of forbidden 
lines, e.g., [Si\,{\sc ii}], [Fe\,{\sc ii}], and [S\,{\sc iii}] lines. Other 
forbidden lines, e.g., [O\,{\sc i}], [C\,{\sc ii}], [N\,{\sc ii}], and [O\,{\sc 
iii}] lines, were identified by Lorenzetti et al. (1999) by the analysis of the 
Long Wavelength Spectrograph (LWS-ISO) spectra of this star. We have also tentatively
identified some features as being caused by H$_2$ molecule, however 
these identifications are very uncertain. McGregor et al. (1988) cited the 
probable presence of CO bands, however we could not identify these features in 
the SWS01 ISO spectrum.

Regarding the dust emission, the ISO-SWS spectrum presents the called 
``dual-dust" chemistry, characterized by the existence of solid state bands of 
C-rich and O-rich dust in the same environment. The C-rich material is 
represented by PAH emission bands (Jourdain de Muizon, d'Hendecourt \& Geballe 1990; Voors 1999). 
On the other hand, beyond 20 $\micron$, the O-rich dust is represented by some features 
caused by crystalline silicates. 

In Table 3, we compile the identifications available from different papers and those ones 
determined by this work, giving a complete view of the most important features 
present in the ISO-SWS spectrum of CD-42$\bmath{\degr}$11721, from 2 to 50\,$\mu$m. 
There, the observed wavelength (Col. 1), the proposed identification (Col. 2), and the reference (Col. 3) for each feature is given.

However, all the spectral description cited above, in this section, probably is not related to the star itself. This happens because ISO and IRAS data cover a large area of the sky around CD-42$\bmath{\degr}$11721 and therefore they are probably contaminated by the known surrounding reflection nebula and/or other stars, since it lies in a crowded region (Habart et al. 2003; Hamaguchi et al. 2005). This contamination is clearly shown by Acke \& van den Ancker (2006) that presented new IR spectra of the 3 and 10\,$\mu$m wavelength regions. These spectra cover an area of only 1$\arcsec$ of diameter around CD-42$\bmath{\degr}$11721, therefore the region close-by this star, and surprisingly, they do not present the strong PAH emission seen in the ISO and IRAS spectra. This fact means that ``dual-dust" chemistry is not related to the object itself.

\begin{figure}
\resizebox{\hsize}{!}{\includegraphics{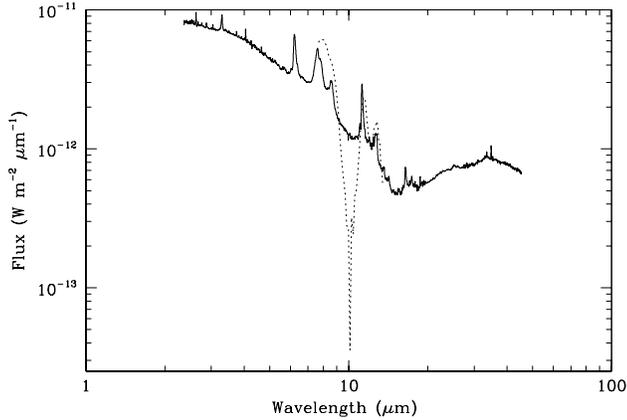}}
\caption{The SWS01 ISO spectrum (solid line) and the IRAS LRS spectrum (dotted line) of CD-42$\bmath{\degr}$11721.}
\label{figure 10}
\end{figure}

\begin{table}
\caption{The identification of the features present in the ISO-SWS spectrum. There Ident. means the proposed feature identifications, enst. is enstatite, diop. is diopside, fost. is fosterite, meth. gr. is related to methil groups, and Uid means unidentified features. References (Ref.): B = Benedittini et al. (1998), V = Voors (1999). It is important to cite that except where noted, otherwise, the identifications are ours. }
\begin{tabular}{ccccccc}
\hline
$\lambda$ (\micron) & Ident. &   Ref. & $\vert$ &  $\lambda$ (\micron) & Ident. &   Ref. \\
\hline
2.39     & Pf (23-5) &    & $\vert$ & 5.40     & Uid          &  \\
2.41     & Pf (22-5) &    & $\vert$  & 5.60     & Uid       &    \\
2.42     & Pf (21-5) &    & $\vert$  & 5.70     & PAH       &  V \\
2.43     & Pf (20-5) &  B & $\vert$  &  6.20     & PAH & V \\
2.45     & Pf (19-5) &    & $\vert$  &  7.60     & PAH & V \\
2.47     & Pf (18-5) &  B & $\vert$  &         & Hu$\beta$ & \\
2.50     & Pf (17-5) &  B & $\vert$  &  7.80     & PAH & V \\
2.53     & Pf (16-5) &  B & $\vert$  & 8.30     & Uid &  V \\
2.56     & Pf (15-5) &  B & $\vert$  & 8.60     & PAH & V \\
2.63     & Br$\beta$      & B & $\vert$  &  11.05     & PAH & V \\
         & Pf (14-5) &  & $\vert$  & 11.23     & PAH & V \\
2.68     & Pf (13-5) &  B & $\vert$  & 12.40     & Hu$\alpha$ & V \\
2.75     & Pf (12-5) &  B & $\vert$  & 12.77     & PAH & V \\
2.87     & Pf (11-5) &  B & $\vert$  & 13.15     & PAH & V \\
3.00     & H$_2$ ?      &    & $\vert$  & 13.39     & Uid &  \\
3.04     & Pf$\epsilon$ & B & $\vert$  & 13.60     & PAH & V \\
3.08     & Uid          &    & $\vert$  & 13.96     & Uid &  \\
3.23     & H$_2$ ?       &    & $\vert$  & 14.20     & PAH & V \\
3.30     & PAH          &  V & $\vert$  & 14.80     & Uid &\\
         & Pf$\delta$ &  & $\vert$  & 15.40     & Uid &   \\
3.40     & meth. gr. &  & $\vert$  & 15.60     & Uid&   \\
3.43     & Uid          &   & $\vert$  & 16.40     & PAH & V \\
3.47     & Uid          &   & $\vert$  & 17.05     & H$_2$ ? &  \\
3.64     & Hu (19-6)  &  & $\vert$  & 17.38     & PAH & V \\
3.70     & Hu (18-6)  & B & $\vert$  & 17.70     & Uid &  \\
3.75     & Hu (17-6)  & B & $\vert$  &  17.86     & [Fe\,{\sc ii}] & V \\
         & Pf$\gamma$ &  & $\vert$  & 18.60     & fost.   &  \\
3.82     & Hu (16-6)  & B & $\vert$  & 18.70     & [S\,{\sc iii}] & V  \\
3.91     & Hu (15-6)  & B & $\vert$  & 19.05     & fost. &   \\
3.97     & Uid          &  & $\vert$  & 19.30     & fost. + enst.  \\
3.99     & Uid          &   & $\vert$  & 25.99     & [Fe\,{\sc ii}] & V  \\
4.02     & Hu (14-6)          & B & $\vert$  & 26.00     & fost., enst. bend  & \\
4.05     & Br$\alpha$ & B & $\vert$  & 29.40     & Uid & V  \\
4.11     & Uid          &   & $\vert$  & 29.60     & Uid &  \\
4.17     & Hu (13-6)          & B & $\vert$  & 31.00     & fost. &   \\
4.19     & H$_2$ ?      &   & $\vert$  & 31.20     & Uid & V \\
4.25     & Uid      &    & $\vert$  & 33.05     & Uid &   \\
4.44     & Hu (12-6) & B & $\vert$  &  33.20     & enst. & V \\
4.65     & Hu (11-6)  & B & $\vert$  & 33.50     & [S\,{\sc iii}]  & V \\
         & Pf$\beta$ &  & $\vert$  & 33.55     & fost. & \\
4.70     & H$_2$ ? &   & $\vert$  & 34.50     & enst. + diop.  &  \\
5.06     & H$_2$ ? &   & $\vert$  & 34.80     & [Si\,{\sc ii}] & V \\
5.13     & Hu$\delta$ & B & $\vert$  & 35.30     & enst.  &  \\
5.23     & PAH          &   & $\vert$  &    & [Fe\,{\sc ii}] & V\\
\hline
\end{tabular}
\label{Table_3}
\end{table}

\section[]{The stellar parameters of CD-42$\bmath{\degr}$11721}\label{parameters}
 
Most of the problems in classifying CD-42$\bmath{\degr}$11721 are caused by
the uncertain stellar parameters. We therefore use firstly our optical high-resolution data
for a tentative determination of the stellar parameters. The details on how
we estimate each parameter are given in this section.

\subsection{The effective temperature}

The effective temperature values for CD-42$\bmath{\degr}$11721 found in the 
literature vary between 12\,300\,K \citep{Hillenbrand} and 31\,600\,K (Cidale et al. 2001). Since in our spectra there is the presence of emission He{\sc i} lines and the absence of He{\sc ii} lines, we can therefore conclude that the effective temperature must be in the range between 13\,000\,K (see Machado \& de Ara\'ujo 2003) and 30\,000\,K.

IUE spectra have been used to constrain the spectral type of CD-42$\bmath{\degr}$11721 to early B (or B0) by Shore et al. (1990). This result should however be taken with caution, because: (i) CD-42$\bmath{\degr}$11721 lies in a crowded region with even some close-by   X-ray sources (see Hamaguchi et al. 2005), (ii) with an elliptical aperture with extensions of $10\arcsec \times 20\arcsec$, IUE observed definitely more than one star, and (iii) the IUE flux is much stronger and clearly offset from our optical spectrum, also indicating that multiple sources have been observed. The IUE spectrum can therefore not be used for a proper temperature determination and the classification as B0 (or early B-type) star is therefore not justified. 

The absence or presence of individual ionization stages of the metals 
identified in our spectrum (see Table\,\ref{Table1}) can be also used for a better 
restriction of the effective temperature. Inspecting the ionization potentials 
for the different ions with observed lines in our spectra, we can state
that CD-42$\bmath{\degr}$11721 shows clear hints for the presence of ions with 
an ionization potential below $\chi \simeq 25$\,eV like e.g. C{\sc iii}, O{\sc 
ii}, N{\sc ii}, Ne{\sc ii}, He{\sc i} recombination lines, and (even though 
very weak) S{\sc iii}. On the other hand, no lines from ions with ionization 
potential higher than $\sim 27$\,eV could be identified. Good tracers for
a higher ionization potential are the lines from Ar{\sc iii}, N{\sc iii}, and Cl{\sc iii} as have been observed e.g. from the compact planetary nebula B[e] 
star Hen 2-90 \citep{Kraus05}. From this narrow range between 25\,eV and 27\,eV, which is just above the He{\sc i} ionization potential, we conclude that the effective temperature can only slightly exceed the value of 13\,000\,K (Machado \& de Araujo 2003). In fact, in order to reproduce the shape of the optical continuum (which for a given extinction is strongly
temperature sensitive, see Sect. 5.4), the effective temperature cannot exceed the value of about 15\,000\,K. We thus conclude that CD-42$\bmath{\degr}$11721 has an effective 
temperature of $14\,000\,K \pm 1\,000\,K$.

\subsection{The distance}

The most controversal parameter of CD-42$\bmath{\degr}$11721 is its 
distance. Literature values range from 136\,pc \citep{Elia} over 160\,pc 
\citep{Hillenbrand} and 400\,pc \citep{deWinter} up to 2\,kpc (McGregor et al. 1988),
2.5\,kpc \citep{Lopes} and even 2.6\,kpc \citep{Shore}. This range in distances 
by about a factor 20 leads to a scatter in stellar luminosity
by a factor 400.
   
For the distance determination from our spectra we make use of the relations 
for the interstellar absorption lines of Na{\sc i} and Ca{\sc ii}-K given
by \citet{Allen}. From the equivalent widths of these lines (see
Table\,\ref{Table1}) we find the following distances:

\begin{table}
\caption{Observed line ratios used for the extinction derivation.}
\begin{tabular}{cccc}
\hline
$n$ & Pa($n$)/H$\epsilon$ & Pa($n$)/H$\delta$ & Pa($n$)/H$\gamma$  \\
\hline
 17 & 1.308 & 0.662 & 0.357 \\
 20 & 0.876 & 0.443 & 0.239 \\
 21 & 0.799 & 0.404 & 0.218 \\
 24 & 0.476 & 0.241 & 0.130 \\
\hline
\end{tabular}
\label{line_ratios}
\end{table}

\begin{itemize}
\item $d = 1.23$\,kpc derived from the Na{\sc i} lines, and
\item $d = 1.06$\,kpc derived from the Ca{\sc ii}-K lines 
\end{itemize}

Both values agree quite well and we will adopt the mean value 
of $d \simeq 1.15$\,kpc, which lies between the very small and very large
distances found in the literature, as a reasonable value. Based on the relations used, we can estimate the error in distance. Taking into account the errors in the measurements of the equivalent widths, on the order of 20 $\%$, the error in distance results in 0.15 kpc. We are aware that this error does not reflect the full error, that might in principle be larger than this, since the distance estimated from the use of interestellar absorption lines contains a systematic error due to the statistical relations used. In our opinion, this parameter remains as the worst estimated, consequently all the conclusions that depend directly on it must be taken with caution.

\subsection{The interstellar extinction}

The extinction towards CD-42$\bmath{\degr}$11721 is believed to be
quite high. Hints for this are provided by the strength of the interstellar
diffuse feature at 4430\,\AA~and the faint IUE SWP spectrum \citep{Shore}.
Literature values for the visual interstellar extinction, $A_{\rm V}$, are 
found to range from 4.2\,mag (Cidale et al. 2001) to 7.1\,mag \citep{deWinter}. 
With such an enormous range of extinction values, it is impossible to draw any 
reliable conclusion from our optical spectra. We therefore attempt to further constrain the extinction.

As a first attempt we tried to use the Pagel method \citep{Pagel}. This method is based on a relation between the observed fluxes of Fe\,{\sc ii} forbidden lines and the reciprocal wavelengths. However, in the case of CD-42$\bmath{\degr}$11721, it could not be employed due to the lack of enough non-blended [Fe\,{\sc ii}] lines in 
our high-resolution spectra. 

We therefore make use of the method suggested by \citet{Kraus07} that is based on the line ratios of the Paschen 
lines over the Balmer lines.
The red edge of our FEROS spectrum shows the presence of well
resolved Paschen lines from Pa(10) up to Pa(40).
In addition, on the blue edge of the spectrum, we observe the Balmer lines
up to H9. The most unaffected line ratio to derive the extinction according to \citet{Kraus07} is the line ratio Pa(10)/H9.
Since the FEROS data are not flux calibrated, we derive the line strengths
by calibrating the data with the Cassegrain spectrum. This spectrum has,
however, a narrower wavelength range than FEROS. We extrapolated the continuum
of the Cassegrain spectrum to the blue and red edges of FEROS, but the lines
in these ranges have a much larger uncertainty and are therefore not so useful
to derive the reddening.
We therefore concentrate on the line ratios Pa($n$)/H$\epsilon$,
Pa($n$)/H$\delta$, Pa($n$)/H$\gamma$ with $n = 15\ldots 25$. The reasons for
these restrictions are the following:

\begin{itemize}
\item Paschen lines from levels higher than $n = 25$ are weak and 
a proper determination of their line strengths becomes difficult.
\item Paschen lines from levels lower than $n = 15$ suffer from the uncertain
continuum extrapolation beyond the red edge of the Cassegrain spectrum.
\item The H9 line is affected by the uncertain continuum extrapolation of
the Cassegrain spectrum which has a rather noisy continuum at the blue edge.
\item The H8 line is not useful since it is blended with a He{\sc i} line.
\end{itemize}

Restricting to unblended, clearly resolved Paschen lines (see Fig.\,\ref{Fig.1})
reduces the lines suitable for a proper extinction determination to Pa(17), 
Pa(20), Pa(21), and Pa(24). For these lines the observed line ratios 
with the Balmer lines H$\epsilon$, H$\delta$, and H$\gamma$ are summarized in 
Table\,\ref{line_ratios}. 

To determine the extinction, we need to know (at least approximately) the
electron temperature in the wind and the mass flux of the star. With an 
effective stellar temperature of about 15\,000\,K and the fact that in a 
stellar wind the Paschen lines form very close to the central star where the 
wind temperature is still rather high, we use the unreddened line ratios for 
the two different electron temperatures, 12\,000\,K and 15\,000\,K.
The mass flux of the star is not known a priori. But since the different line 
ratios should result in the same (or a very similar) extinction value, we can 
exclude the very high mass fluxes (i.e. $F_{\rm m} \ge 
10^{-4}$\,g\,s$^{-1}$cm$^{-2}$) in the tables of \citet{Kraus07}, 
because they result in a large spread in extinction values.
 
\begin{table}
\caption{Mean extinction values, $A_{\rm V}$ in magnitudes, from the line 
ratios calculated for different mass fluxes and electron temperatures. The 
Paschen lines used to calculate the mean values are Pa(17), Pa(20), Pa(21), 
and Pa(24).}
\begin{tabular}{llcccc}
\hline
 & & \multicolumn{4}{c}{\textrm \protect{$F_{\rm m}$ [g\,s$^{-1}$cm$^{-2}$]}} \\
 & $T_{\rm e}$ [K] & $10^{-5}$ & $10^{-6}$ & $10^{-7}$ & $10^{-8}$ \\
\hline
Pa($n$)/H$\epsilon$ & 12\,000 & 5.065 & 4.775 & 4.476 & 4.398 \\
Pa($n$)/H$\delta$   & 12\,000 & 4.960 & 4.571 & 4.298 & 4.276 \\
Pa($n$)/H$\gamma$   & 12\,000 & 5.153 & 4.688 & 4.493 & 4.521 \\[0.3em]
Pa($n$)/H$\epsilon$ & 15\,000 & 4.995 & 4.697 & 4.409 & 4.344 \\
Pa($n$)/H$\delta$   & 15\,000 & 4.855 & 4.464 & 4.208 & 4.197 \\
Pa($n$)/H$\gamma$   & 15\,000 & 4.984 & 4.535 & 4.354 & 4.393 \\
\hline
$<A_{\rm V}>$ & 12\,000 & 5.059 & 4.678 & 4.422 & 4.398 \\
$<A_{\rm V}>$ & 15\,000 & 4.945 & 4.565 & 4.324 & 4.311 \\
\hline
\end{tabular}
\label{av_mean}
\end{table}

To pin down the extinction, we first calculate a mean value of the different 
line ratios found for each mass flux in the range $F_{\rm m} = 10^{-5}\ldots
10^{-8}$\,g\,s$^{-1}$cm$^{-2}$ and for the two different electron temperatures.
These mean values are given in Table\,\ref{av_mean}. From this table it is 
clear already, that the visual extinction must be in the range from 4.3\,mag to
5.0\,mag, which is in rough agreement with the values of 4.2\,mag and 4.3\,mag 
found by Cidale et al. (2001) and \citet{Hillenbrand}, and the value of about 5\,mag
given by McGregor et al. (1988).

To restrict the extinction even further, we next calculate the recombination
line luminosities of the Paschen lines, using the model parameters for $T_{\rm 
e}$ and $F_{\rm m}$, redden them with 
the corresponding extinction value found, and compare them with the observed 
line luminosities under the assumption of a distance of 1.15\,kpc.
Such a comparison can give only a rough result, because the line
luminosities, different from the line ratios, do depend on individual
stellar parameters like the stellar radius and mass loss rate. 
Assuming spherical symmetry, which should result in a lower limit of the
real mass flux of the star, and varying the stellar parameters in a reasonable 
range, we found that even with this rough method
we can exclude the mass fluxes lower than $10^{-6}$\,g\,s$^{-1}$cm$^{-2}$.
This means, that the extinction towards CD-42$\bmath{\degr}$11721 is
on the order of $4.8\pm 0.2$\,mag.

\subsection{Stellar radius and luminosity}

The stellar radius of CD-42$\bmath{\degr}$11721 can be found by fitting
the Cassegrain spectrum with classical Kurucz model atmospheres \citep{Kurucz}.
Since the effective temperature of the star is in the range 13\,000 to 15\,000\,K, we fitted Kurucz model atmospheres for stars with solar metallicity and different $\log g$ values, reddened with $4.8\pm 0.2$\,mag interstellar extinction. From these fittings it turned out that no  reliable fit was found for $A_{\rm V} = 5.0$\,mag, while for the values of $A_{\rm V} = 4.6 - 4.8$, especially the models with low to mid surface gravity, fitted best (Fig. 7). From the fitting parameter which is simply $R_*^{2}/d^{2}$ and for a distance 
of $1.15\pm 0.15$\,kpc we can derive the stellar radius. This has been done for all the best fit model atmospheres. The resulting stellar radius taking into account the error in distance is $R_{*} = 17.3\pm 0.6$\,R$_{\odot}$. We want to stress that this is an upper limit for the stellar 
radius, because the optical continuum might contain some contributions from 
free-free and free-bound emission generated in the wind. Nevertheless, the 
contribution of the wind can only be of minor impact because the photometric 
data in the near infrared do not show influences of free-free emission (see Sect. 6.1). We 
therefore regard the derived $\sim 17$\,R$_{\odot}$ as a reasonable value for the 
stellar radius. Since we know the radius and the effective temperature, we could find a stellar luminosity of $L_{*} = (1.0\pm 0.3)\times 10^{4}$\,L$_{\odot}$.

To finish our optical analysis, we summarize in Table 6 the stellar parameters of CD-42$\bmath{\degr}$11721 derived within this section.

\begin{figure}
\includegraphics[width=84mm]{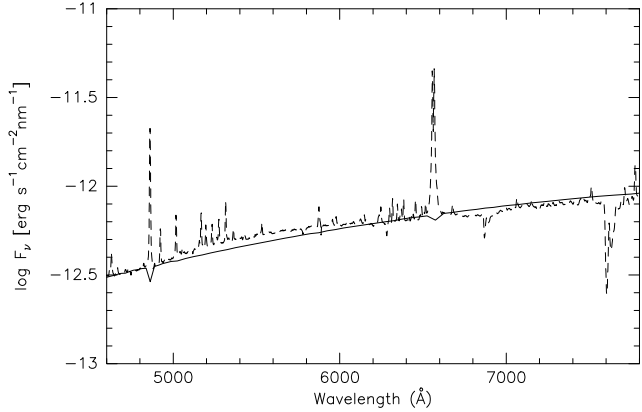}
 \caption{Fitting of the observed optical continuum (dashed line) with Kurucz 
model atmospheres (solid lines, \citealt{Kurucz}). Shown is one of the best fits achieved for a solar metalicity star with $T_{\rm eff} = 14\,000$\,K, $\log g = 3.0$, and a foreground extinction of $A_{\rm V} = 4.8$\,mag.}
\label{spec_fit}
\end{figure}

\begin{table}
\caption{Stellar parameters for CD-42$\bmath{\degr}$11721 found from our
high-resolution optical spectra as described in Sect.\,\ref{parameters}. 
The mass and $\log g$ values have been derived under the assumption that
CD-42$\bmath{\degr}$11721 is a post-main sequence object. The values in the second row represent the error in each measurement.
}
\begin{tabular}{@{}ccccccc}
\hline
$T_{\rm eff}$ & $R_{*}$ & $L_{*}$ & $M_{\rm initial}$ & $\log g$ & $d$ & 
   $A_{\rm V}$ \\
$ $ [K] & [R$_{\odot}$] & [L$_{\odot}$] & [M$_{\odot}$] &  & [kpc] & [mag]  \\ 
\hline
14\,000 & 17.3 & $1.0\times 10^{4}$ & 8--10 & $\sim 3$ & 1.15 & 4.8 \\
\hline
$\pm 1000$ & $\pm 0.6$ & $\pm 0.3\times 10^{4}$ &  & & $\pm 0.15$ & $\pm 0.2$ \\ 
\hline
\end{tabular}
\label{stel_param}
\end{table}

\section{Modelling of the SED of CD-42$\bmath{\degr}$11721}

To construct a proper SED for CD-42$\bmath{\degr}$11721, we collected all photometric data available in the literature and listed them in Table 7. There, the observed wavelengths (Col. 1), the measured fluxes (Col. 2), the aperture diameters (Col. 3), and the references (Col. 4) are given. As can be seen in Fig. 8, considering all data and different apertures, the SED is double peaked, similar to those HAeBe stars called "group I" (Meeus et al. 2001), though differently than other objects of this group, its first peak is shifted to higher wavelengths, around 2 $\micron$. However, as cited in Sect. 4, the data from observations with large apertures are contaminated by the reflection nebula and other sources and do not come from the star itself. Based on this, we have decided to consider in our modelling only data with aperture diameters equal or smaller than 15$\arcsec$. It is important to cite that the most confident results would be obtained from the modelling of a SED considering only photometric data using apertures around 1$\arcsec$ or 2$\arcsec$, however there is not enough available data to build a reliable SED for this object.

\begin{table}
\caption{Photometric data of CD-42$\bmath{\degr}$11721 available in the literature. Refs: 1 = de Winter et al. (2001); 2 = Herbst (1975); 3 = Berrilli et al. (1992); 4 = Elia et al. (2005); 5 = Mannings (1994); 6 = Hillenbrand et al. (1992); 7 = Henning et al. (1994); 8 = Skinner, Brown \& Stewart (1993); ISO-SWS = Acke \& van den Ancker (2004); MSX = Egan et al. (1999); IRAS = http://irsa.ipac.caltech.edu/IRASdocs/iras$\_$mission.html; 2MASS = http://www.ipac.caltech.edu/2mass/releases/allsky/doc/sec1$\_$6b.html}
\begin{tabular}{@{}ccccc}
\hline
 Wavelength  & Flux  & Aperture Diameter$^*$ & References  \\
 $ $ [$\micron$] & [$W m^{-2} \micron^{-1}$] & [$arcsec$] & \\
\hline
0.36 & 4.67$\times 10^{-13}$ & 15.0  & 1$^{**}$ \\
0.36 & 3.86$\times 10^{-13}$  &  9 - 13 &   2 \\
0.44 & 6.23$\times 10^{-13}$  & 15.0  & 1$^{**}$  \\
0.44 & 5.44$\times 10^{-13}$  &  9 - 13 &  2 \\
0.56 & 1.16$\times 10^{-12}$  & 21.5  &  1$^{**}$ \\
0.56 & 1.07$\times 10^{-12}$  & 15.0  &  1$^{**}$ \\
0.56 & 9.22$\times 10^{-13}$  &  9 - 13 &   2 \\
0.64 & 2.17$\times 10^{-12}$  & 15.0  &  1$^{**}$  \\
0.79 & 2.66$\times 10^{-12}$  & 15.0  &  1$^{**}$ \\
1.23 & 3.69$\times 10^{-12}$   & 15.0  &  3 \\
1.25 & 4.40$\times 10^{-12}$   & 13.0  &  1$^{**}$ \\
1.25 & 3.92$\times 10^{-12}$  &  8.0 &  2MASS \\
1.63 & 5.11$\times 10^{-12}$   & 15.0 & 3  \\
1.65 & 5.37$\times 10^{-12}$  &  13.0 & 1$^{**}$  \\
1.65 & 4.81$\times 10^{-12}$  &  8.0  &  2MASS  \\
2.19 & 6.60$\times 10^{-12}$  &  15.0  & 3   \\
2.20 & 7.11$\times 10^{-12}$   & 13.0  & 1$^{**}$    \\
2.20 & 6.33$\times 10^{-12}$  & 8.0  & 2MASS  \\
3.30 & 6.50$\times 10^{-12}$  & 14 x 20 & ISO-SWS \\
3.40 & 6.49$\times 10^{-12}$  &  14 x 20 & ISO-SWS  \\
3.45 & 8.31$\times 10^{-12}$  &  13.0  & 1$^{**}$  \\
3.50 & 6.29$\times 10^{-12}$   & 14 x 20 & ISO-SWS  \\
3.79 & 5.83$\times 10^{-12}$  & 15.0 &  3 \\
4.29 & 6.86$\times 10^{-12}$    & 18.3  & MSX  \\
4.35 & 6.89$\times 10^{-12}$  & 18.3 & MSX  \\
4.64 & 4.87$\times 10^{-12}$  & 15.0 & 3  \\
4.80 & 4.37$\times 10^{-12}$  & 13.0 &  1$^{**}$  \\
6.20 & 3.23$\times 10^{-12}$  & 14 x 20 &  ISO-SWS  \\
7.70 & 2.24$\times 10^{-12}$  & 14 x 20 &  ISO-SWS  \\
8.28 & 3.26$\times 10^{-12}$  & 18.3  & MSX \\
8.38 & 1.73$\times 10^{-12}$ & 15.0 & 3 \\
8.60 & 1.77$\times 10^{-12}$  & 14 x 20  & ISO-SWS  \\
9.69 & 9.06$\times 10^{-13}$  & 15.0 & 3 \\
9.70 & 1.26$\times 10^{-12}$ & 14 x 20 & ISO-SWS  \\
11.00 &  1.14$\times 10^{-12}$  & 14 x 20 & ISO-SWS  \\
12.00 & 1.98$\times 10^{-12}$  & 45.0$\times$270.0 & IRAS \\
12.00 &  9.26$\times 10^{-13}$ &   80.0  &   4 \\
12.13 & 1.54$\times 10^{-12}$  & 18.3 & MSX  \\
12.89 &  4.90$\times 10^{-13}$ & 15.0 & 3 \\
14.65 & 6.47$\times 10^{-13}$  & 18.3 & MSX  \\
21.34 & 7.84$\times 10^{-13}$  & 18.3 & MSX  \\
25.00 & 1.21$\times 10^{-12}$  & 45.0$\times$276.0 &  IRAS  \\
25.00 &  6.68$\times 10^{-13}$  &  80.0 &  4 \\
60.00 & 1.60$\times 10^{-12}$  & 90.0$\times$282.0  &  IRAS  \\
60.00 & 1.31$\times 10^{-12}$  & 80.0  &  4 \\
100.00  & 6.63$\times 10^{-13}$  & 180.0$\times$300.0 & IRAS \\
100.00 & 5.12$\times 10^{-13}$  & 80.0  &  4 \\
350.00 & 2.32$\times 10^{-16}$   & 18.5  & 5 \\
450.00 & 8.02$\times 10^{-17}$  & 18.3  & 5 \\
600.00 & 1.11$\times 10^{-17}$  & 17.5  & 5  \\
750.00 & 4.74$\times 10^{-18}$  & 18.0  & 5  \\
800.00 & 5.06$\times 10^{-18}$  & 17.0  & 5  \\
850.00 & 2.86$\times 10^{-18}$  & 18.0  & 5 \\
1100.00  & 6.20$\times 10^{-19}$  & 18.8 & 5  \\
1300.00 & 4.58$\times 10^{-19}$   & 19.8  & 5  \\
1300.00 &  4.79$\times 10^{-19}$  & 28.0  &  6   \\
\hline
\end{tabular}
\label{phot_param}
\end{table}

\begin{table}
\contcaption{ }
\begin{tabular}{@{}ccccc}
\hline
 Wavelength  & Flux  & Aperture Diameter$^*$ & References  \\
 $ $ [$\micron$] & [$W m^{-2} \micron^{-1}$] & [$arcsec$] & \\
\hline
1300.00 &  7.54$\times 10^{-19}$ &  23.0  &  7  \\
36000.00 & 3.06$\times 10^{-24}$ &  2.5 &  8   \\ 
\hline
\end{tabular}
\small{$ $ * = except by IRAS and ISO-SWS, that have rectangular apertures. ** = Mean values of observations with same aperture and from the same reference. }
\end{table}

\begin{figure}
\includegraphics[width=80mm]{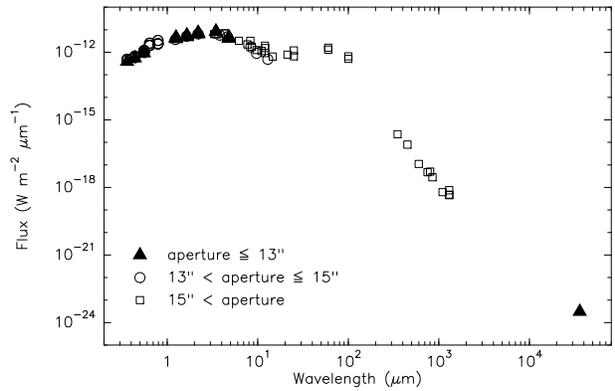}
 \caption{The SED of CD-42$\bmath{\degr}$11721, considering all data with different apertures presented in Table 7.}
 \label{figure 8}
\end{figure}

In this section, we present the modelling of the SED of CD-42$\bmath{\degr}$11721, using three 
diferent codes, that solve the problem of radiative transfer in a circumstellar 
dust medium, considering three different scenarios, e.g. a spherically symmetric 
geometry, a passive flared disc and an outflowing disc wind. We aim to determine which scenario describes better the SED of this star, assuming the set of stellar parameters obtained from our optical analysis. Our codes that represent the flared disc and the outflowing disc wind scenario are presented here for the first time.

\subsection{Free-free and free-bound emission}

For the modeling of the SED of CD-42$\bmath{\degr}$11721, our first step was to determine the
contribution of the free-free and free-bound emission (hereafter ff-fb emission)
to the total continuum. The ff-fb continuum can be calculated by accounting for
the observed 3.6\,cm emission that was obtained with an aperture size of 2.5"
(see Table 7) and that can be ascribed purely to the ionized wind material.
Since we do not know exactly the wind geometry and the possible inclination
under which the system of CD-42$\bmath{\degr}$11721 is seen, we calculated the ff-fb continuum
assuming an isothermal ($T_e = 10000 K$) spherically symmetric wind, with an outer edge of $1500 AU$. The mass loss rate needed to fit the 3.6\,cm emission ($\dot{M} = 2.86\times 10^{-6} M_{\odot}yr^{-1}$) must be considered as a lower limit. The terminal velocity, which is proportional to the escape velocity on the stellar surface, was calculated from
the stellar parameters given in Table\,6 and it is $515 km s^{-1}$. It turns out, that the contribution
of the ff-fb continuum is negligible in the optical and near-IR spectral range (see Fig. 9).

\subsection{Spherically symmetric envelope} 

Our optical results revealed that the circumstellar scenario of 
CD-42$\bmath{\degr}$11721 should probably be composed by a non spherical 
geometry. However, for sake of completeness, we modeled the SED of this star considering also a circumstellar spherically symmetric dust envelope.

Our numerical treatment of radiative transfer in a
spherical envelope was described in Lorenz-Martins \& Pompeia (2000).
The Monte Carlo scheme is used for representing the propagation of
radiative energy photon by photon. For each interaction between a
``photon'' and a grain, a fraction of the energy is stored
(absorption) and the remaining part is scattered according to the
scattering diagram. Two species of grains are used and
they interact with each other. The stellar radiation
leads to a first distribution of dust temperature and the thermal
radiation from grains is simulated, giving after several iterations
the equilibrium temperature. Here two species of grains are
considered both with a numerical density decreasing as $r^{-2}$, which
corresponds to an expansion at constant velocity. The grains can be
present at same or at different distances of the central star. The following 
physical quantities are required in a first guess and are subsequently fixed by 
fitting the infrared/near infrared flux: the effective temperature of the 
central star, $T_{eff}$; grain sizes and refractive indices for
all wavelengths; the extinction opacity, $\tau_{ext}$ at $\lambda$ = 9 $\mu$m;  
the inner radii of the dust envelope (say, R$_{Sil}$, inner radius for amorphous  
silicate grains; R$_{Crys}$, inner radius for crystalline silicate grains, or 
R$_{A.C.}$ amorphous carbon grains), and R$_o$ the outer radius of the dust 
envelope.

The models give not only the spectral repartition of the total
flux but also of its different components (direct, scattered, thermal
for each species of grain), as well as the angular distribution of
these radiation fields and the radial dependence of dust temperature.
The absorption and scattering efficiencies, as well as the albedo for
the grains were calculated by us using the Mie theory and the optical
constants given by Rouleau \& Martin (1991), David \& P\'egouri\'e (1995)
and Jaeger et al. (1998).

\begin{table}
\caption{Results of the modelling considering a spherical symmetric envelope. The temperatures are given in $K$. More details are in the text.}
      \begin{flushleft}
      \begin{tabular}{cccccc}     
      \hline\noalign{\smallskip} 
Model & T$_{eff}$ & T$_{A.C.}$ & T$_{Sil}$ & T$_{Crys_1}$ & T$_{Crys_2}$ \\
      \noalign{\smallskip}            
      \hline\noalign{\smallskip} 
 A  & 10000 & 564  & 364 & -- & -- \\
 B  & 15000 & 781  & 570 & -- & -- \\
 C  & 25000 & 1038 & 880 & -- & -- \\
 D  & 10000 & -- & -- & 346 & 327 \\
         \hline
      \end{tabular}
      \end{flushleft}
\end{table}

Several models, considering different sets of parameters, including 
those derived by our optical analysis, were calculated by us. The results of four of them 
are presented in Table 8. There, the name of each model (Col. 1) is followed by $T_{eff}$ (in $K$, Col. 2) and the temperatures (also in $K$) of the hottest grains (Cols. 3, 4, 5 and 6). Other parameters are kept constant in these models. These are $\tau_{ext} = 3.0$ at $\lambda = 9$, R$_{Sil}$ = R$_{Crys}$ = R$_{A.C.}$ = $200 R_*$ and R$_o$ = $5000 R_*$. The grains sizes are also fixed at $200 \AA$ and the ratio of amorphous carbon over amorphous silicate grains is set to $0.50$ and the ratio of crystalline silicate grains needed in model D is also set to $0.5$. 

As can be seen in Fig. 9, the total SEDs (including the stellar and dust contributions) obtained by us, assuming a spherical geometry for the dust envelope, could not model properly the SED of 
CD-42$\bmath{\degr}$11721, especially the optical region. 
The model A and D differ one from the other only by the dust species (see Table 8). In model A we have used amorphous silicates and amorphous carbon grains 
simultaneously, while in model D we have considered only 
crystalline silicates. The other two models (B and C) differ from the model A only by $T_{eff}$ as can be seen in Table 8. These models present very similar fitting, even considering quite different set of parameters.  

\begin{figure}
\resizebox{\hsize}{!}{\includegraphics{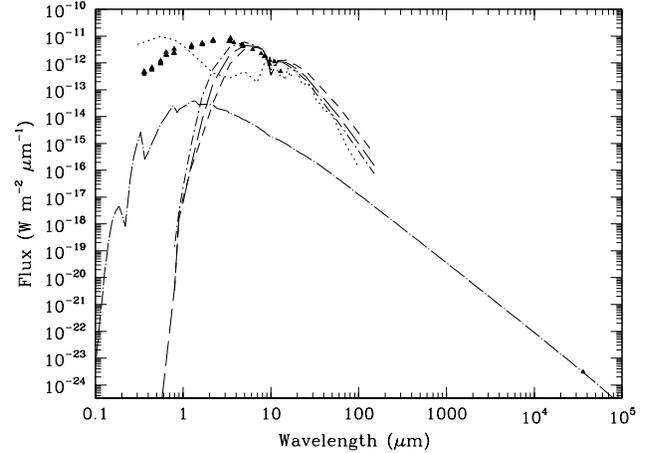}}
\caption{Different models reproducing the SED of CD-42$\bmath{\degr}$11721, considering a spherical symmetric envelope. We can see the model A (dashed line), model B (long-dashed line), model C (dot-short dashed line) and model D (dotted line). A detailed explanation of these models can be seen in the text and also in Table 8. The triangles represent the photometric data obtained in the literature. The dot-long-dashed line is the ff-fb emission that fits the 3.6\,cm data point. Its contribution, as can be seen, to the IR part of the SED is negligible.}
\label{figure 9}
\end{figure}

\subsection{Equatorial disc}

Based on our optical analysis, a more complex circumstellar geometry, most plausibly a disc scenario, seems to be indeed necessary to account for the observed SED of CD-42$\bmath{\degr}$11721.

The presence of a disc seems to be the better answer to explain the 
spectral 
characteristics of not only B[e] supergiants (Zickgraf et al. 1985), but also of 
HAeBe stars. The 
discs around young stars are detected with millimeter interferometry (Mannings 
\& Sargent 1997, 2000; Fuente et al. 2004), and in some cases, they are visible in optical 
images of the 
stars, either via scattered light
or in silhouette against a background light source (Burrows et al. 1996; Grady 
et al. 2000; McCaughrean \& O'Dell 1996).

Due to the absence of images and also interferometric observations for 
CD-42$\bmath{\degr}$11721, we have decided to reproduce its SED considering two 
different types of discs, one typical of HAeBe stars and another one, typical of 
B[e]sg stars.

\subsubsection{Passive flared disc}

Considering CD-42$\bmath{\degr}$11721 as an HAeBe star, we need to assume the 
existence of a passive equatorial disc, where passive means that the disc is 
heated purely by stellar irradiation.

Specifically for CD-42$\bmath{\degr}$11721, we need to explain the location of 
the hot dust 
that might be responsible for the peak seen in its SED. We are 
assuming here that the hot dust is located at the surface of the circumstellar 
disc. For the surface to become much hotter than the interior, it must be 
inclined with respect to the infalling stellar radiation, i.e. the disc must be 
flared. Such a flaring can happen naturally when the disc is in hydrostatic 
equilibrium in vertical direction with a radial temperature distribution that 
falls off more slowly than $r^{-1}$ (Kenyon \& Hartmann 1987). 

The flared disc is proposed to explain the spectral characteristics of young 
pre-main sequence stars of the ``group I" (Meeus et al. 2001; Dullemond \& 
Dominik 2004; Acke et al. 2005). The geometry of this kind of disc also explains the presence of a more intense flux in the far-IR, seen in HAeBe stars of the ``group I", since the most distant regions in the disc surface receive a 
higher amount of stellar radiation (they see a larger solid angle of the star). This kind of disc can also explain the 
presence of the ``dual-dust" chemistry, since the PAH molecule emission would be formed in the surface layers of the disc and the silicate bands in the inner regions of it, closer to the star (Meeus et al. 2001; Dullemond \& Dominik 
2004). However, as cited in the Sect. 4, in the case of CD-42$\bmath{\degr}$11721, this mixed chemistry probably does not belong to the circumstellar medium of this star. 

Considering a flared disc scenario, we developped an analytical two-layer model that contains an optically thick ($\tau_{\rm V}^{\rm mid} \gg 1$) mid-layer which is isothermal in $z$-direction. The mid-layer is sandwiched by
two top-layers which are marginally optically thick at visual wavelengths but
still optically thin at IR wavelengths, i.e. $\tau_{\rm V}^{\rm
top} \ge 1 \quad$ and $\quad \tau_{\rm IR}^{\rm top} \le 1$. Consequently,
the infalling stellar light is completely absorbed within the top-layer. The
dust particles re-radiate the energy at IR wavelengths, so half
the redistributed energy leaves the disc into space, and the other half
penetrates the mid-layer and heats it.
As a third distinct region we define the disc photosphere which encloses the
uppermost part of the top-layer and whose location is defined by the parameter
$h$ (see below).

Further, we make the following assumptions: the disc is in hydrostatic
equilibrium in $z$-direction, gas and dust are well mixed throughout the disc,
and in the mid-layer gas (g) and dust (d) are in thermal equilibrium at the
same temperature $T_{\rm g} = T_{\rm d} = T_{\rm mid}$.
The emission and absorption of the gas component is ignored.

In addition, our (gas)
disc extends down to the stellar surface with no inner hole. The existence of
such a hole would lead to a puffed-up inner rim and self-shadowing effects of
the disc (see Dullemond, Dominik \& Natta 2001 and Dullemond 2002).

For a razor-thin passive disc (see Fig. 10) around a star with 
effective temperature $T_{*}$ and radius $R_{*}$ the monochromatic flux 
entering the disc perpendicular to the surface at distance $r$ from the star is
\begin{eqnarray}\label{f_real}
f^{\perp}_{\nu} & = & \int\limits_{0}^{\pi}\int
       \limits_{0}^{\theta_{\rm max}} B_{\nu}(T_{*})\sin^{2}\theta\sin\phi\,
        {\rm d}\theta\,{\rm d}\phi\nonumber \\
 & = & B_{\nu}(T_{*})\left[\arcsin\frac{R_{*}}{r} -
\frac{R_{*}}{r}\sqrt{1-\left(\frac{R_{*}}{r}\right)^{2}}~\right]
\end{eqnarray}
with
\begin{equation}
\theta_{\rm max} = \arccos\left(\sqrt{1-\left(\frac{R_{*}}{r}
\right)^{2}}\right)
\end{equation}
We assume that the star emits a Planck spectrum.
At distances large compared with the stellar radius, this relation reduces to
\begin{equation}\label{f_approx}
f^{\perp}_{\nu} \stackrel{r\gg R_{*}}{\longrightarrow} B_{\nu}(T_{*})
\frac{2}{3}\left(\frac{R_{*}}{r}\right)^{3}
\end{equation}
The flux can be parametrized in terms of the solid angle $\Omega$
under which the star is seen from a disc surface element at distance $r$ and 
the so-called grazing
angle $\alpha_{\rm gr}$, i.e. the mean angle of incidence of the stellar flux
\begin{equation}\label{alfa_def}
f^{\perp}_{\nu} = \alpha_{\rm gr} \Omega B_{\nu}(T_{*})
\end{equation}
The grazing angle of a razor-thin disc is therefore
\begin{equation}\label{alfa_razor}
\alpha_{\rm gr}^{\rm razor}  =  \frac{1}{\Omega}\left[\arcsin\left(
\frac{R_{*}}{r}\right) - \frac{R_{*}}{r}\,\sqrt{1-\left(\frac{R_{*}}
{r}\right)^{2}}~\right]
\end{equation}
which for large distances reduces to the handy formula first introduced by 
Chiang \& Goldreich (1997)
\begin{equation}\label{alfa_razor_red}
\alpha_{\rm gr}^{\rm razor} \stackrel{r\gg R_{*}}{\longrightarrow}~\frac{4}
{3\pi}\frac{R_{*}}{r} \simeq 0.4 \frac{R_{*}}{r}
\end{equation}
This relation also holds for a wedge-shaped disc as long as its opening angle 
is small.

In a flared disc the surface is curved and the grazing angle increases with 
distance from the star (see the solid lines in Fig. 11). 
Therefore, the flared disc intercepts more stellar light and is heated more 
efficiently. The grazing angle of the flared disc is by the amount $\alpha_{\rm 
gr}^{\rm flare}$ higher than for a flat disc (Fig. 11). This 
flaring term of the grazing angle can be computed from
\begin{equation}\label{alfa_h}
\alpha_{\rm gr}^{\rm flare}  =  \arctan \frac{{\rm d}h}{{\rm d}r} -
\arctan\frac{h}{r}\simeq r \frac{\rm d}{{\rm d}r}
\left(\frac{h}{r}\right)
\end{equation}
and the total grazing angle becomes simply 
\begin{equation}
\alpha_{\rm gr} = \alpha_{\rm gr}^{\rm razor} + \alpha_{\rm gr}^{\rm flare}
\end{equation}
With this new grazing angle, which is a function of $h(r)$, i.e. the bottom
of the photosphere or half the thickness of the disc, we can calculate the flux 
penetrating the flared disc in $z$-direction.

\begin{figure}
\resizebox{\hsize}{!}{\includegraphics{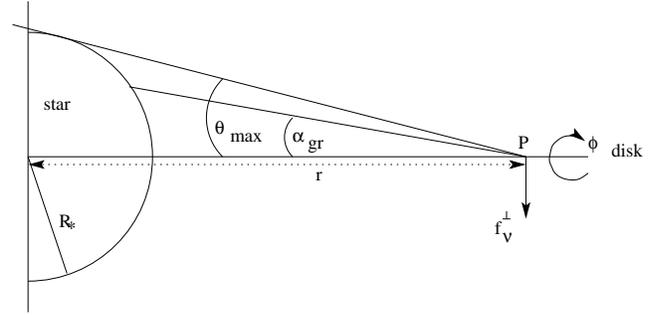}}
\caption{Sketch of stellar illumination of a razor thin disc. The grazing angle
$\alpha_{\rm gr}$ (see text for details) is also indicated.}
\label{Fig. 10}
\end{figure}

\begin{figure}
\resizebox{\hsize}{!}{\includegraphics{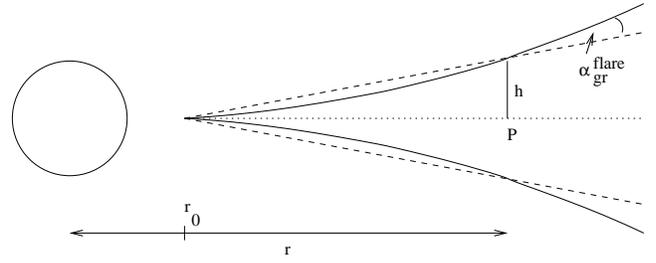}}
\caption{Sketch of a flared disc. The flaring term of the grazing angle is
indicated. The inner radius of the dust disc is $r_{0}$.}
\label{Fig. 11}
\end{figure}

The photosphere encloses the uppermost part of the top-layer. Its onset is 
described
by the parameter $h$, which is defined as the height $z$ above the mid-plane 
where
the visual optical depth along the direction of the infalling stellar light, 
i.e.
along the grazing angle, equals 1. This leads to a vertical optical depth of
\begin{equation}\label{tau_perp}
\tau^{\perp}_{\rm V} = \sin \alpha_{\rm gr}\,.
\end{equation}
Since the disc is assumed to be in hydrostatic equilibrium, $\tau^{\perp}_{\rm 
V}$
can be calculated
\begin{equation}\label{tau2}
\tau_{\rm V}^{\perp} = \kappa_{\rm V}\rho_{0}(r)\int\limits_{h}^{\infty}e^{-
\frac{z^{2}}{2H^{2}}}\,dz
\end{equation}
where $\rho_{0}$ is the density in the mid-plane and $H$ is the scale height of 
the 
disc given by
\begin{equation}\label{scale}
H = \sqrt{\frac{kT}{GM_{*}\mu m_{\rm H}}}~r^{3/2}
\end{equation}
From equaling Eqs.\,(\ref{tau_perp}) and (\ref{tau2}) $h$ can be determined, 
but it is a function of temperature and grazing angle.

The downwards and upwards directed fluxes, $F^{\downarrow}$ and $F^{\uparrow}$,
are in equilibrium throughout the passive disc.
We can calculate these two fluxes explicitely
at the boundary between the isothermal mid-layer and the top-layer.
All infalling stellar radiation is reprocessed within the top-layer.
We assume that half the incident stellar flux is re-radiated into space and
the other half penetrates the mid-layer. The downwards directed flux becomes
\begin{equation}\label{fdown}
F^{\downarrow} = \frac{1}{2}\,f^{\perp} = \frac{1}{2}\int \alpha_{\rm gr}
\Omega B_{\nu}(T_{*})\,d\nu
\end{equation}
and the flux leaving the mid-layer in upward direction is 
\begin{equation}\label{fup}
F^{\uparrow} = 2\pi\int\!\!\!\int B_{\nu}(T_{\rm mid})\left(
1-e^{-\tau_{\nu}/\mu}\right)\mu\,d\mu\,d\nu
\end{equation}
where $T_{\rm mid}$ is the isothermal temperature of the mid-layer, and we set 
$\mu = \cos\theta$ with $\theta$ as the angle measured from the $z$-axis. The 
visual optical depth of the mid-layer in vertical direction can be written in 
the form
\begin{equation}
\tau_{\rm V}^{\rm mid}(r) = \tau_{\rm V}^{\rm mid}(r_{0})~\left(\frac{r}{r_{0}}
\right)^{-s}
\end{equation}
with the visual optical depth at the inner edge, $\tau_{\rm V}^{\rm
mid}(r_{0})$, and the exponent $s$ as free parameters, and the optical depth
at frequency $\nu$ is simply
\begin{equation}\label{tau_nu}
\tau_{\nu} = \tau_{\rm V}~\frac{\sigma^{\rm ext}_{\nu}}
{\sigma^{\rm ext}_{\rm V}}
\end{equation}
where $\sigma^{\rm ext}$ is the dust extinction cross section.
The temperature of the mid-layer, $T_{\rm mid}$, follows from equaling 
Eqs.\,(\ref{fdown}) and (\ref{fup}). This temperature is, however, a function 
of the grazing angle.

We have now three important parameters, $\alpha_{\rm gr}, h,$ and $T_{\rm mid}$,
but none of them can be computed independently, instead they must be calculated
iteratively.

We have to specify the visual optical depth in the top-layer in $z$-direction,
$\tau^{\rm top}_{\rm V}$.  It should be high enough so that first, the infalling
stellar radiation is completely absorbed and second, the dust grains at the
bottom of the top--layer reach a temperature close to that of the mid-layer in
order to guarantee a smooth transition.  On the other hand, $\tau^{\rm top}_{\rm
V}$ must be small enough to allow the dust emission, which occurs at infrared
wavelengths, to escape the top-layer.

The radiation transfer equation of a plane-parallel slab is
\begin{equation}\label{strahl}
I_{\nu}(\mu,\tau_{\nu}) = I_{0}\,e^{\tau_{\nu}/\mu} - \int
S_{\nu}(t)\,e^{-(t-\tau_{\nu})/\mu}\,\frac{dt}{\mu}
\end{equation}
with the source function $S_{\nu}$, the incident intensity $I_{0}$ and
$\mu = \cos\theta$. The radiation transfer equation is split
into up-streams, $I^{+}_{\nu}$, that penetrate the top-layer from the
mid-layer, and down-streams, $I^{-}_{\nu}$, that cross the top-layer
starting from the surface. The incident intensity is either the
stellar radiation for downwards directed streams (but only for angles
$\theta$ under which the star can be seen, else it is zero), or the
emission of the mid-layer, $B_\nu(T_{\rm mid})(1-e^{-\tau_\nu^{\rm mid}/\mu})$,
for the upwards directed streams.

With the help of the Feautrier parameters
\begin{eqnarray}
u_{\nu} & = & \frac{1}{2}\,(I^{+}_{\nu} + I^{-}_{\nu}) \\
\upsilon_{\nu} & = & \frac{1}{2}\,(I^{+}_{\nu} - I^{-}_{\nu})
\end{eqnarray}
the mean intensity $J_{\nu}$ becomes
\begin{equation}
J_{\nu} = J_{\nu}(\tau_{\nu}) = \int u_{\nu}(\mu,\tau_{\nu})\,d\mu
\end{equation}
which is needed to calculate the source function and the emission of the
grains. The source function itself determines the up- and down-streams of
the intensity which leads to an iterative calculation of the radiation field. 

Finally, the temperature of the grains at each location in the
top-layer follows from balancing their absorption from the surrounding
radiation field, $J_{\nu}$, and their emission at their equilibrium
temperature, $T_{\rm d}$.

In our code, we are adopting a mixture of spherical silicates and amorphous 
carbon grains which follow the MRN grain size distribution (Mathis, Rumpl \& Nordsieck 1977). 
The particle sizes range from $a_{\rm min} = 75$\,\AA~to $a_{\rm max} \simeq 
1\,\mu$m. The absorption 
efficiencies of the dust particles are calculated with the Mie theory for 
spherical particles and scattering of the stellar light within the top-layer is
included. 

For the modelling of CD-42$\bmath{\degr}$11721, assuming a flared disc scenario, we have used the physical parameters obtained by our optical analysis (extinction, effective temperature, distance, and luminosity) and the ff-fb emission (see Sect. 6.1). We keep the top-layer optical depth fixed ($\tau^{\rm top}_{\rm V} = 2$) and use a stellar mass of $8-10 M_{\odot}$ (see Sect. 7).

Considering this set of input parameters, the code calculates the location of the inner edge of the dust disc, which is defined by the dust evaporation temperature and may vary for different input values, especially of the optical depth of the mid-layer (i.e. the surface density distribution). Having fixed the inner edge of the disc, the disc structure and temperature distribution in vertical direction is calculated. Of special importance is thereby that the grain temperatures of the different grain sizes and dust components adopt to the isothermal temperature of the mid-layer. The dust SED is then calculated by integrating the emerging dust emission over the disc size.

The most serious boundary condition in our calculations is the request of an optically thick mid-layer, justifying the application of a flared disc. Intending to fulfill this condition and to reproduce the observed dust emission, we need to play mainly with the outer edge of the dust disc ($r_{out}$), the mid-layer optical depth at the inner edge of the dust disc ($\tau^{\rm mid}_{\rm V}(r_{0})$), and the exponent ($s$). Assuming values of $s$ higher than 1.5, we obtain that only very small discs ($r_{out} \leq 25 AU$) can keep the mid-layer optically thick, however in these cases, the dust disc emission is lower than the observed one. On the other hand, assuming $s = 1$ or $1.5$, the density distribution is less steep and we can satisfy the mid-layer optical depth condition for larger discs and consequently the dust emission is better reproduced. We can also get good results assuming $\tau^{\rm mid}_{\rm V}(r_{0})$ in the range of $2.5\times 10^2$ and $4\times 10^3$. Concerning the $r_{out}$, it is limited by the mid-layer optically depth condition. 

The total SED is formed by the stellar, ff-fb and dust disc emission contributions. The stellar contribution is determined by the stellar parameters, especially the effective temperature and also the foreground extinction, obtained by our optical analysis. In the panel (a) of Fig. 12 we present our best result. It clearly shows deviations compared to the observations, especially the near-IR part that cannot be well fitted by our calculations. In this model, we are considering a fixed density profile with $s = 1$, $r_{0}$ = $0.5 AU$, $r_{out}$ = $60 AU$, and $\tau^{\rm mid}_{\rm V}(r_{0}) = 2.5\times 10^2$. On the other hand, an increase of the mid-layer optical depth (i.e. an increase in surface density) has caused an increase in far-IR flux and consequently, it cannot be adjusted to the observations. 

It is important to cite that the inclination angle of the disc, seen by the observer, is also an input parameter of our code and we have found that there is no difference in our model results, as long as we consider discs with inclination angles smaller than $70\degr$ ($0\degr$ means pole-on). For larger inclination angles it is impossible to get a reasonable fit at all.

\begin{figure}
\resizebox{\hsize}{!}{\includegraphics{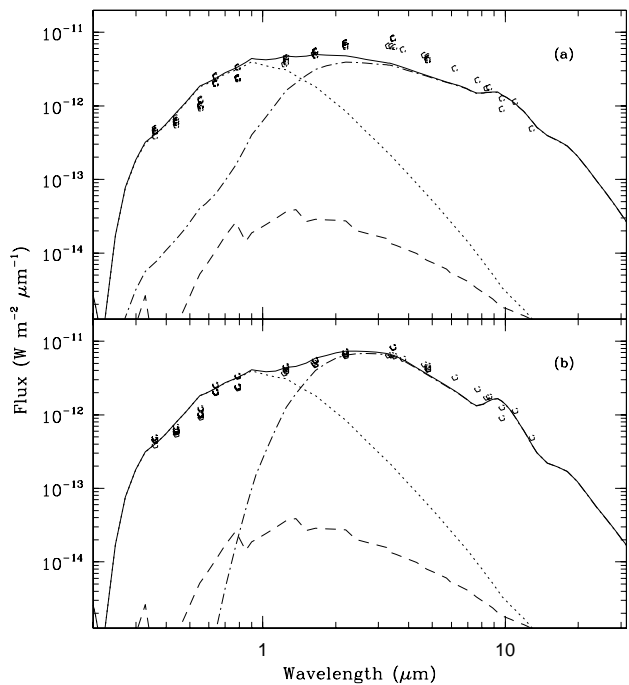}}
\caption{Modelling the SED of CD-42$\bmath{\degr}$11721, considering in the panel (a) a flared disc scenario and assuming a less steep density profile ($s = 1$). As can be seen, the observations cannot be well reproduced, because the near-IR emission is clearly underestimated. The panel (b) show our best model considering the presence of an outflowing disc wind. The models were obtained assuming a disc seen with an inclination angle from 0$\degr$ (pole-on) to $70\degr$. The circles represent the photometric data obtained in the literature. The dotted line represents the stellar contribution; the dashed line, the ff-fb emission; the dot-dashed line, the dust disc contribution and the solid line, the total emission. }
\label{figure 12}
\end{figure}

\subsubsection{Outflowing disc wind}

We have alternatively supposed the disc around CD-42$\bmath{\degr}$11721 to be an outflowing disc wind, typically ascribed to B[e] supergiants, where the dust would be uniformly 
distributed in it. Our 
code, for simplification, assumes that the disc is isothermal 
in z-direction, with a constant outflow velocity and a surface density described by: 
\begin{equation}
\Sigma(r) = \int\limits_{-h}^{h}\rho(r,z) dz = 2 \int\limits_{0}^{h}\rho(r,z) dz
\label{surf}
\end{equation}
where $h$ is here the height of the disc at distance $r$, considering a 
constant opening angle for it, and $\rho(r,z)$ is the density distribution, being 
derived according to the mass 
continuity equation:

\begin{equation}
\rho(r,z) = \frac{\dot{M}}{4\pi (r^{2} + z^{2}) v_{\infty}}
\end{equation}
Inserting this equation into Eq. (20), it follows that  

\begin{equation}
\Sigma(r) = \Sigma_{0} \left(\frac{r}{r_0}\right)^{-1} = \Sigma_{0} 
\left(\frac{r_0}{r}\right)
\end{equation}
with 
\begin{equation}
\Sigma_{0} = \frac{\dot{M}\alpha}{4\pi r_0 v_{\infty}} 
\label{sigma}
\end{equation}
and $\alpha$ is here the opening angle of the outflowing disc and $r_0$ is the dust sublimation radius. Eq. (22) represents the surface 
density of the gas component. To convert it into the surface density of the dust 
component, we consider the gas-to-dust
ratio, that is a factor 100. For the temperature distribution we assume a power law of the form

\begin{equation}
T(r) = T_{0} \left(\frac{r}{r_{0}}\right)^q
\end{equation}
where $T_{0}$ is the dust sublimation temperature, fixed at 1500 K.

We use the same grain size distribution and grain composition as for the flared disc. The model output is then the integrated flux of the star plus disc system, with the stellar parameters taken from our optical analysis, and reddened with the foreground extinction. The inclination angle of the star-disc system is a free input parameter.

The exponent of the temperature distribution ($q$, see Eq. 24) is found to be the parameter with most severe influence on the resulting SED. We have obtained good results assuming different values of $q$ in the range from $0$ to $-2$. For values higher than $-0.4$, we could get good results, assuming quite low disc mass loss rates ($\dot{M}_{disc} \leq 10^{-7} M_{\odot}yr^{-1}$), however the main problem in this case is that the temperature continues still high far away from the central star, what is not expected. On the other hand, using $q \leq -0.5$, it is necessary very high disc mass loss rates ($\dot{M}_{disc} \geq 10^{-4} M_{\odot}yr^{-1}$) or even large disc opening angles ($\alpha \geq 40\degr$) to reproduce the observed SED. 

The panel (b) of Fig. 12 shows our best result, considering the following parameters: $\alpha = 20\degr$, $r_{out} \geq 20 AU$, $\dot{M}_{disc} = 8.7\times 10^{-5} M_{\odot}yr^{-1}$ and $q = -0.4$. This result seems for us better than those ones obtained assuming a flared disc scenario. Regarding to the outer egde of the disc ($r_{out}$), we can get the same good results, assuming $r_{out}$ with values from $20 AU$ to $8000 AU$ (which corresponds to approximately the size of the aperture considered). This means the disc is not optically thick anymore just after few $AUs$ of distance from the star and the contribution from the optically thin region is negligible for the SED. This conclusion is in agreement with the fact that we can model the SED assuming a disc seen pole-on or with inclination angles lower than 70$\degr$, meaning that whole emission region is seen under these angles. Another interesting point is that our models are obtained assuming values of disc mass loss rate that are in perfect agreement with those ones proposed for discs of B[e] supergiants. Another parameter that also influences our models is the terminal velocity of the disc wind (a decrease of $v_\infty$ causes an increase of the dust disc emission), however based on our optical observations (see Table 1), we decide to keep it fixed, $v_\infty = v_{exp} = 60 km s^{-1}$.

\subsection{Discussion: model results}

In the case of CD-42$\bmath{\degr}$11721 several works have tentatively described its 
SED using different codes and approaches.  Hillenbrand et al. (1992) suggested  
that the SED of this star shows no evidence of a spherically symmetric envelope. 
According to 
these authors, the SED could be described by a flat, optically thick massive 
accretion disc. These discs present an optically thin region within the inner 
few stellar radii. These optically thin "holes" would imply the absence of warm 
dust 
which would contribute with strong excess of emission in the near infrared 
region and that is seen in the SED of CD-42$\bmath{\degr}$11721. However they 
did not treat the radiative transfer problem and their fit are not good, 
even at 3 $\mu$m. Natta et al. (1993) calculated models to 
CD-42$\bmath{\degr}$11721 taking the radiative transfer into account. Their 
model was solved considering a spherically symmetric dust envelope which would
reprocess the radiation of the central source. The difference in this work 
is that they adopted a central source composed by a star and a circumstellar 
disc.  In addition, they modified the dust properties at $\lambda$ $>$ 100 
$\mu$m in order to match the observed submillimeter fluxes.  Anyway, the model 
fits relatively well the data, except at longer wavelengths than 100 $\mu$m. 
 
Radiative transfer models were also calculated by Henning et al. (1994) with a 
spherical geometry for the dust envelope. Their model fits well the overall SED, 
including longer wavelengths than 100 $\mu$m.  They suggested that the main part 
of the radiation comes from a spherical dust configuration. More recently, Elia 
et al. (2004) modeled the continuum emission from the circumstellar environment 
of some sources using the same code as Henning et al. (1994). Their models were 
fitted to ISO data, considering both, gas and dust arranged in circumstellar 
spherical shells with temperature and density profiles described by power laws, 
T $\approx$ r$^{-q}$ and n $\approx$ r$^{-p}$, respectivelly. They also 
considered a disc geommetry but they did not fit longer wavelengths. 

In our study we have firstly analyzed the photometric data available in the literature and based on the aperture sizes used, we have selected only those points that could be related to star (or the close-by regions) and not contaminated by other sources and/or nebula. Using our code which treats the radiative transfer via the Monte Carlo 
method for a spherical circumstellar medium, we could not to 
reproduce the SED of CD-42$\bmath{\degr}$11721. Even using some physical 
parameters cited by Henning et al. (1994), especially the envelope inner radius, 
the effective temperature, and density law (n $\approx$ r$^{-1}$) we have 
obtained unrealistic high grain temperatures (about 7000 K), meaning that the 
dust could not be located so close to the star. We are aware of the fact that 
there are several differences between both codes, however we can point out three 
of them. Firstly, their model is dependent of the distance of the object (luminosity) and they are assuming a higher value than us for CD-42$\bmath{\degr}$11721 ($d = 2 kpc$), and secondly 
they considered two different density 
distributions inside of the envelope, 
working at different distances from the star. However not only the structure of 
the circumstellar region determines the shape of a SED, but also the nature of 
the grains itself. Henning et al. (1994) adopted a fluffy dust model to describe 
dust particles in such a circumstellar medium.  No small particles were 
considered. That is the third difference between our model and Henning's one, 
and it may be the most important. Usually a fluffy dust model is used to 
describe the coagulation process in dense cores of molecular clouds (Ossenkopf 
1993) and young pre-main-sequence stars, but it is not applicable to hot stars. Therefore, we conclude that a spherically symmetric scenario is ruled out.

Based on our optical analysis and the results described previously, we believe that a non-spherical circumstellar envelope, possibly an 
equatorial disc, seems to be necessary to explain the SED of the 
CD-42$\bmath{\degr}$11721. Considering a flared disc scenario, we cannot model well its SED, especially the near- and the far-IR regions. On the other hand, considering an outflowing disc wind, we could get better results. It is important to cite that a better and even more conclusive modelling will be obtained only when more data using small apertures, especially from far-IR region, are available.

\section{The nature of CD-42$\bmath{\degr}$11721}

\begin{figure}
\includegraphics[width=80mm]{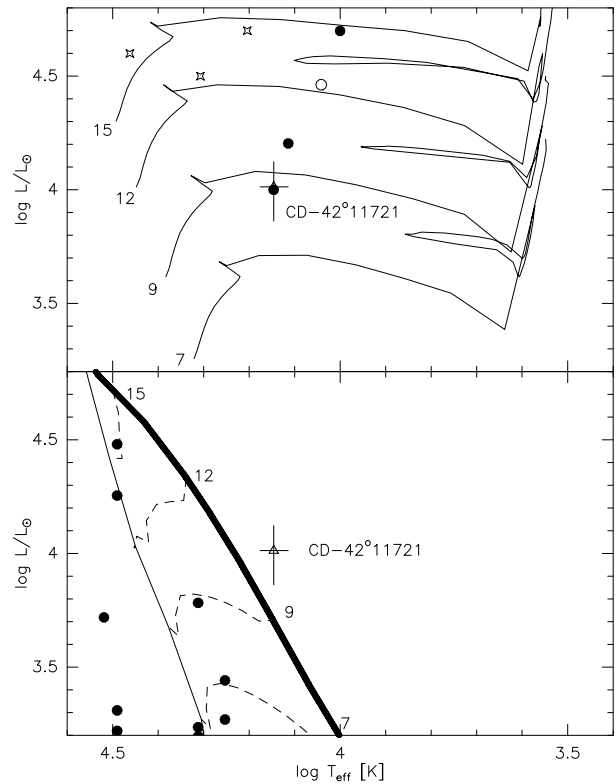}
 \caption{Top panel: The HR diagram for B[e] supergiants in the Large Magellanic Cloud
(filled circles), the Small Magellanic Cloud (open circles), and the
Milky Way (asterisks), data from Lamers et al. (1998) and references therein.
Also plotted are evolutionary tracks from \citet{Schaller} for non-rotating 
stars with solar metallicity. The location of CD-42$\bmath{\degr}$11721 
is in agreement of being a post-main sequence object having evolved from a  
progenitor star with an initial mass of about 8--10\,M$_{\odot}$. In addition, it
lies in the region of well known B[e] supergiants.
Bottom panel: The HR diagram for galactic HAeBe stars
\citep[filled circles,][]{Berrilli}. The solid line defines the zero-age main
sequence, the thick solid line defines the birthline. CD-42$\bmath{\degr}$11721
is plotted as the triangle. It is located above the birthline.}
 \label{hrd}
\end{figure}

As we presented in this paper, CD-42$\bmath{\degr}$11721 is a very curious object presenting the B[e] 
phenomenon. Its evolutionary stage is quite doubtful, being considered either a 
young pre-main sequence star (HAeB[e]) or an evolved post-main sequence 
supergiant (B[e] supergiant) and all this doubt is caused by the completely absence of reliable physical parameters for this star.  

In our work, we have firstly presented an optical analysis of this object, that is based on new low- and high-resolution (highest than previously published) optical spectra of
CD-42$\bmath{\degr}$11721. Especially the high-resolution spectrum reveals a huge number of emission lines from
both, permitted and forbidden transitions. Many of these lines have been
detected for the first time. We created a spectral atlas containing the
equivalent widths of all measured lines, as well as a profile classification.
Line profiles are found to be single-peaked, double-peaked, or (especially
lines of Fe{\sc ii}) even multiple-peaked. The presence of atoms mostly neutral or singly ionized and the variety in line profiles speak in favour of a 
non-spherical density distribution within the circumstellar medium close to the
central star. The existence of a circumstellar disc
is suggested especially by the existence of numerous emission lines
from neutral elements. 

Using these optical data, we could estimate the stellar parameters of CD-42$\bmath{\degr}$11721. We plotted the effective temperature and luminosity obtained by us, into the HR diagram (top panel of Fig.\,\ref{hrd}) together with known galactic and MC B[e] supergiants, taken from the summary table given in Lamers et al. (1998) and references therein. Also included are evolutionary tracks taken from \citet{Schaller} for non-rotating stars with solar metallicity. The position of CD-42$\bmath{\degr}$11721 falls into the region of B[e] supergiants ($L_{*} \ge 10^{4}$\,L$_{\odot}$, Lamers et al. 1998) and even coincides with the position of Hen S59, a LMC B[e] supergiant. While all to date known galactic B[e] supergiants have $L_{*} \ge 3 \times 10^{4}$\,L$_{\odot}$, there are some low luminosity objects ($10^{4} \le L_{*}(L_{\odot}) \le 3 \times 10^{4}$) in the Magellanic Clouds (Gummersbach, Zickgraf \& Wolf 1995), for which CD-42$\bmath{\degr}$11721 might be the first galactic low luminosity counterpart, having evolved from a progenitor star with an initial mass of 8--10\,M$_{\odot}$. With such a mass, the surface gravity becomes $\log g \simeq 3$. 

To test a possible pre-main sequence nature, we plotted the HR diagram for galactic HAeBe stars taken from
\citet{Berrilli} together with the birthline, the zero-age main sequence, and
pre-main sequence evolutionary tracks from \citet{Bernasconi}. While the
galactic HAeBe stars lie well below the birthline, this is not the case for
CD-42$\bmath{\degr}$11721, making its classification as a HAeBe star
questionable (bottom panel of Fig.\,\ref{hrd}).

Intending to improve this discussion, we have also made an IR analysis of this star. As shown in the Sect. 4, Acke \& van den Ancker (2006) showed that the most detached characteristic of the ISO and IRAS spectra, the very intense PAH features, actually do not seem to be related to the star itself, but to an envelope or a surrounding nebula. This fact is very important, because the presence of C-rich material implies in a C-rich photosphere (Waters et al., 1998) that seems hard to be explained in a supergiant scenario whose spectrum is dominated by several H recombination lines, even knowing that PAH emission was already identified in LBV stars (Voors 1999). In the Sect. 6, we have also presented that our best results for the modelling of CD-42$\bmath{\degr}$11721 were obtained considering an outflowing disc wind, typical of B[e] supergiant stars.  

Summarizing, based on our optical and IR analysis (even considering that our errors could be underestimated) and the modelling of the SED, a B[e] supergiant nature seems to us to be a good scenario to explain the observed characteristics of 
CD-42$\bmath{\degr}$11721. However, a young or even pre-main sequence nature for this curious object could not definitely be excluded by our qualitative analysis.

\section*{Acknowledgments}

M.B.F. acknowledges financial support from \emph{CNPq} (Post-doc position - 
150170/2004-1), Utrecht University, LKBF and NOVA foundations (The Netherlands). 
M.B.F. also acknowledges Rens Waters and Henny Lamers for the pleasant and 
fruitful discussion about the nature of CD-42$\bmath{\degr}$11721 in Vlieland (The Netherlands), during the "Workshop on Stars with the B[e] Phenomenon". 
M.K. was supported by GA AV under grant number KJB300030701.

\appendix

\section[]{The optical line list of CD-42$\bmath{\degr}$11721}\label{tab}

\begin{table}
\caption{Catalogue of lines identified in the FEROS spectrum of 
CD-42$\bmath{\degr}$11721. Listed are the observed wavelength, equivalent
width, description of the line profile (S = single-peaked, D = double-peaked,
M = multiple-peaked, A = absorption, blend = superposition of several lines),
and the line identification with the laboratory wavelength.}
\begin{tabular}{cccc}
\hline
Wavelength (\AA)  &   W($\lambda$)  & Line Profile & Identification \\
\hline
3835.3  &   1.64   &   S  & H9 3835.4 \\
3855.6  &   1.46  &  S & Si\,{\sc ii} (m1) 3856.0 \\
3889.0  &   3.56   &  S   & He\,{\sc i} (m2) 3888.7 \\
        &          &      & H8 3889.1 \\
3933.5  &   0.34  &  A & Ca\,{\sc ii} (m1) 3933.7\\
3968.4  &   0.13  &  A & Ca\,{\sc ii} (m1) 3968.5\\
3969.4  &   2.33   &  S & H$\epsilon$ 3970.1 \\
4068.3  &    0.22   & S  &  [S\,{\sc ii}] (m1F) 4068.6 \\
4101.7  &   4.61   & S &  H$\delta$ 4101.7 \\
4173.3 &   0.51  &  S  & Fe\,{\sc ii} (m27) 4173.5 \\
4178.5 &   0.75  &  S  & Fe\,{\sc ii} (m21) 4177.7 \\
4233.0 &   1.70  &  S  & Fe\,{\sc ii} (m27) 4233.2 \\
4243.4 &   0.50  &  S  & [Fe\,{\sc ii}] (m21F) 4244.0 \\
4258.0 &   0.22  &  S  & Fe\,{\sc ii} (m28) 4258.2 \\
4276.4 &   0.14  &  S  & [Fe\,{\sc ii}] (m21F) 4276.8 \\
4287.0 &   0.57  &  S  & [Fe\,{\sc ii}] (m7F) 4287.4 \\
4290.1 &   0.13  &  S  & Ti\,{\sc ii} (m41) 4290.2 \\
        &       &       & Ne\,{\sc ii} (m57) 4290.4 \\
4296.3 &   0.36  &  D  & Fe\,{\sc ii} (m28) 4296.6 \\
4303.0 &   0.48  &  S  & Fe\,{\sc ii} (m27) 4303.2 \\
4340.5  &   8.76   & S &  H$\gamma$ 4340.5 \\
4351.5 &   1.24  &  S  & Fe\,{\sc ii} (m27) 4351.8 \\
4358.8 &   0.46  &  S  & [Fe\,{\sc ii}] (m7F) 4359.3 \\
4384.7 &   1.31  &  D  & Mg\,{\sc ii} 4384.6 \\
4413.5  &    0.62   &  S  &  [Fe\,{\sc ii}] (m7F) 4413.8 \\
4416.4  &    2.15   &   S &  [Fe\,{\sc ii}] (m6F) 4414.5 \\
        &          &    & [Fe\,{\sc ii}] (m6F) 4416.3 \\
        &          &    & Fe\,{\sc ii} (m27) 4416.8 \\
4451.7  &    0.30   & S &  [Fe\,{\sc ii}] (m32F) 4452.1 \\
4457.7  &    0.18  &  S & [Fe\,{\sc ii}] (m6F) 4458.0 \\
4472.4  &    0.71   & S &  He\,{\sc i} (m14) 4471.7 \\
4481.1 &    0.98   & S &  Mg\,{\sc ii} (m4) 4481.3 \\
4489.0  &    0.50  &  D & [Fe\,{\sc ii}] (m6F) 4488.8 \\
4491.2 &   0.63  &  D  & Fe\,{\sc ii} (m37) 4491.4 \\
4508.1  &    0.80   &  blend & Fe\,{\sc ii} (m38) 4508.3 \\
        &      &   & Fe\,{\sc ii} [m6F] 4509.6 \\
4515.1  &    1.01   &  M & [Fe\,{\sc ii}] (m6F) 4514.9 \\
4520.0  &    1.05   &  D & Fe\,{\sc ii} (m37) 4520.2 \\
4522.4       & 1.08   &  D   &   Fe\,{\sc ii} (m38) 4522.6 \\
4533.9       & 0.25   &  S   &   Fe\,{\sc ii} (m37) 4534.2 \\
4541.3       & 0.36   &  S   &   Fe\,{\sc ii} (m38) 4541.5 \\
4549.3  &    2.07  &  D &  Fe\,{\sc ii} (m38) 4549.5 \\
4555.6  &    1.14  &  M &  Fe\,{\sc ii} (m37) 4555.9 \\
4558.3  &    0.36  &  D &  Fe\,{\sc ii} (m20) 4558.6 \\
4576.1  &    0.29  &  D &  Uid \\
4583.5  &    2.71   & blend &  Fe\,{\sc ii} (m37) 4582.8 \\
        &          &   &  Fe\,{\sc ii} (m38) 4583.8 \\
        &          &   &  Fe\,{\sc ii} (m26) 4584.0 \\
4588.0 &    0.18   & S &  Cr\,{\sc ii} (m44) 4588.2 \\
        &          &   &  Cr\,{\sc ii} (m16) 4588.4 \\
4618.5  &   0.17  &  S  & C\,{\sc ii} 4618.9  \\
4620.3 &    0.32   & S &  Fe\,{\sc ii} (m38) 4620.5 \\
        &          &   &  N\,{\sc ii} (m5) 4621.4 \\
4629.1 &    2.03   & D &  Fe\,{\sc ii} (m37) 4629.3 \\
4634.5 &    0.37   & S &  Fe\,{\sc ii} (m25) 4634.6 \\
4639.4 &    0.11   & S &  O\,{\sc ii} (m1) 4638.9 \\
4657.0 &    0.24   & S &  Fe\,{\sc ii} (m43) 4657.0 \\
4666.5 &    0.39   & D &  Fe\,{\sc ii} (m26) 4665.8 \\
4727.7  &   0.19   &  S & [Fe\,{\sc ii}] (m4F) 4728.1 \\
4731.2  &   0.40    &  D &  Fe\,{\sc ii} (m43) 4731.4 \\
\hline
\end{tabular}
\label{Table1}
\end{table}

\begin{table}
\contcaption{ }
\begin{tabular}{cccc}
\hline
\hline
Wavelength (\AA)  &   W($\lambda$) & Line Profile &  Identification \\
\hline
4814.2  &   0.28    &  S   & [Fe\,{\sc ii}] (m20F) 4814.6 \\
4823.9 &  0.25    & D  &   Cr\,{\sc ii} (m30) 4824.1 \\
        &       &       &  S\,{\sc ii} (m52) 4824.1 \\
4847.9 &    0.22   & S &  Fe\,{\sc ii} (m30) 4847.6 \\
4861.4  & 30.41   &  D & H$\beta$ 4861.3 \\
4889.3  &   0.24   &  S &  [Fe\,{\sc ii}] (m4F) 4889.6 \\
        &      &   &  [Fe\,{\sc ii}] (m3F) 4889.7 \\
4905.0  &   0.16    & S &  [Fe\,{\sc ii}] (m20F) 4905.4 \\
4923.7  &    3.63  & M &  Fe\,{\sc ii} (m42) 4923.9 \\
5018.1 &    4.76   & D &   Fe\,{\sc ii} (m42) 5018.4 \\
5041.5 &    0.49   &  S &   Si\,{\sc ii} (m5) 5041.1 \\
5056.1  &    0.64   & S &  Si\,{\sc ii} (m5) 5056.4 \\
5100.5  &    0.15   & S &   Fe\,{\sc ii} 5101.0 \\
5111.1  &    0.15   & S &   [Fe\,{\sc ii}] (m19F) 5111.6 \\
5145.6  &    0.25   & S &   C\,{\sc ii} (m16) 5145.2 \\
        &           &   &   Fe\,{\sc ii} (m35) 5146.1 \\
5149.1  &    0.08   & S &   Fe\,{\sc ii} 5149.5 \\
5158.3  &    0.55   & M &   [Fe\,{\sc ii}] (m18F) 5158.0 \\
5163.6  &    0.09   & S &   [Fe\,{\sc ii}] (m35F) 5163.9 \\
5168.9  &    3.87   & D &  Fe\,{\sc ii} (m42) 5169.0  \\
5183.4  &    0.09   & S &   Mg\,{\sc i} (m2) 5183.6 \\
5197.4  &   1.98    &  D &   Fe\,{\sc ii} (m49) 5197.6 \\
5203.4  &    0.10   & S & Uid \\
5227.1  &    0.16   & S &  Fe\,{\sc iii} 5227.5 \\
5234.5  &    2.41   & D &  Fe\,{\sc ii} (m49) 5234.6 \\
5237.0  &    1.01   & S & Uid \\
5248.5  &    0.25   & S &  Fe\,{\sc ii} 5248.0 \\
5254.6  &    0.30   & D &  Fe\,{\sc ii} (m49) 5254.9 \\
5256.8  &    0.09   & S &  [Cr\,{\sc ii}] (m13F) 5256.0 \\
5261.5  &    0.38   & D &   [Fe\,{\sc ii}] (m19F) 5261.6 \\
5264.5  &    0.38   & S &  Mg\,{\sc ii} 5264.3 \\
        &           &   &  Fe\,{\sc ii} (m48) 5264.8 \\
5268.5  &    0.11   & S &  [Fe\,{\sc iii}] (m1F) 5270.4 \\
5273.0  &    0.21   & S &  C\,{\sc iii} 5272.6 \\
5273.1  &    0.41   & S &  [Fe\,{\sc ii}] (m18F) 5273.4 \\
5275.7  &    1.85   & D &  Fe\,{\sc ii} (m49) 5276.0 \\
5283.8  &    0.57   & D &  [Fe\,{\sc ii}] (m35F) 5283.1 \\
5291.6  &    0.10   & S &  Fe\,{\sc iii} 5291.8 \\
5296.5  &    0.06   & S &  [Fe\,{\sc ii}] (m19F) 5296.8 \\
5298.7  &    0.08   & S &  Uid \\
5316.3  &    3.99   & D &   Fe\,{\sc ii} (m49) 5316.6 \\
5325.3  &    0.24   & D &   Fe\,{\sc ii} (m49) 5325.6 \\
5333.4  &    0.20   & S &   [Fe\,{\sc ii}] (m19F) 5333.7 \\
5346.6  &    0.21   & S &   Fe\,{\sc ii} (m49) 5346.6 \\
        &           &   &   [Fe\,{\sc ii}] (m18F) 5347.7 \\
5362.6  &    1.20   & blend &   [Fe\,{\sc ii}] (m17F) 5362.1  \\
        &           &   &   Fe\,{\sc ii} (m48) 5362.9 \\
5376.0  &    0.16   & S &   [Fe\,{\sc ii}] (m19F) 5376.5 \\
5402.0  &    0.13  & S &   Mg\,{\sc ii} (m24) 5401.1 \\
5413.2  &    0.21   & S &   [Fe\,{\sc iii}] (m1F) 5412.0 \\
5425.0  &    0.35  &  D &  Fe\,{\sc ii} (m49) 5425.3 \\
5432.7  &    0.22   & S &  Fe\,{\sc ii} (m55) 5432.9 \\
        &          &  & [Fe\,{\sc ii}] [(m18F) 5433.2 \\
5466.6  &    0.25  &  S &  Fe\,{\sc ii} 5466.0 \\
5502.8  &    0.13   & S & Uid \\
5510.7  &    0.11   & S &  Cr\,{\sc ii} (m23) 5510.7 \\
5527.0  &    0.19   & S &  [Fe\,{\sc ii}] (m17F) 5527.3  \\
        &       &       & [Fe\,{\sc ii}] (m34F) 5527.6 \\
5529.5  &    0.22   & S & Uid \\
5534.5  &    1.00  &  D &  Fe\,{\sc ii} (m55) 5534.9 \\
5543.8  &    0.08   & S & Uid \\
5576.7  &    0.84  &  S &  [O\,{\sc i}] (m3) 5577.3 \\
5746.5  &    0.10   & S &  [Fe\,{\sc ii}] (m34F) 5747.0  \\
5754.2  &   0.07    & S &   [N\,{\sc ii}] (m3F) 5754.8  \\
\hline
\end{tabular}
\end{table}

\begin{table}
\contcaption{ }
\begin{tabular}{cccc}
\hline
\hline
Wavelength (\AA)  &   W($\lambda$)  & Line Profile & Identification \\
\hline
5780.4  &   0.44    & A &   DIB  \\
5791.7 &   0.08    & S &   C\,{\sc i} (m18) 5793.5 \\
5797.0  &   0.16    & A &   DIB  \\
5835.1 &  0.07    &  S & Fe\,{\sc ii} (m58) 5835.4 \\
5849.7  &   0.05    & A &   DIB  \\
5876.0 &  2.00    &  S & He\,{\sc i} (m11) 5875.6  \\
5889.7  &  0.65    &  A & Na\,{\sc i} (m1) 5890.0  \\
5895.7  &  0.58    &  A & Na\,{\sc i} (m1) 5895.9  \\
5957.3  &   0.50    & S &   Si\,{\sc ii} (m4) 5957.6 \\
5978.9  &   0.85    & S &  Fe\,{\sc iii} (m117) 5978.9  \\
        &       &       &  Si\,{\sc ii} (m4) 5979.0  \\
5991.0  &   0.27    & M &  Fe\,{\sc ii} (m46) 5991.4  \\
6040.3 &   0.13    & S &  Uid \\
6045.9  &   0.21    & S &  S\,{\sc i} (m10) 6046.0 \\
        &       &       &  O\,{\sc ii} 6046.3 \\
        &       &       &  O\,{\sc ii} 6046.5 \\
6083.7  &   0.13    & S &  Fe\,{\sc ii} (m46) 6084.1  \\
6103.2  &   0.06    & S &  Fe\,{\sc ii} (m200) 6103.5  \\
6125.2  &   0.30    & S &   Mn\,{\sc ii} (m13) 6125.9 \\
6130.1  &   0.32    & S &   Mn\,{\sc ii} (m13) 6129.0 \\
        &       &    &  Fe\,{\sc ii} (m46) 6129.7 \\
6148.2  &   0.72   & M &   Fe\,{\sc ii} (m74) 6147.7 \\
6158.0  &   0.50    & S &  O\,{\sc i} (m10) 6156.0 \\
        &       &       &  O\,{\sc i} (m10) 6156.8 \\
        &       &       &  O\,{\sc i} (m10) 6158.2 \\
6160.8  &   0.14    & S &  Fe\,{\sc ii} (m161) 6160.8 \\
6172.7 &   0.15    & S &   N\,{\sc ii} (m36) 6173.4 \\
6195.8  &   0.06    & A &   DIB  \\
6202.9  &   0.09    & A &   DIB  \\
6233.2  &   0.25    & M &   Fe\,{\sc ii} 6233.5 \\
6238.3  &   0.49    & D &  Fe\,{\sc ii} (m74) 6238.4 \\
6247.8  &   1.40    & M &  Fe\,{\sc ii} (m74) 6247.6 \\
6291.5  &   0.15    & S &   Uid  \\
6299.8  &    1.20  & D &  [O\,{\sc i}] (m1F) 6300.3 \\
6317.8  &    2.17  & D &  Fe\,{\sc ii} 6318.0  \\
6338.0  &   0.13    & S &   Uid  \\
6347.0  &    1.73   & S &   Si\,{\sc ii} (m2) 6347.1 \\
6356.8 &   0.16    & S &   Uid  \\
6363.3  &    0.43   & D &   [O\,{\sc i}] (m1F) 6363.8 \\
6371.1  &   1.39    & S &   Fe\,{\sc ii} (m40) 6369.5 \\
        &          &  &  Si\,{\sc ii} (m2) 6371.4 \\
6384.1  &    2.30   & M &   Fe\,{\sc ii} 6383.8 \\
6416.6  &   0.47    & D &   Fe\,{\sc ii} (m74) 6416.9 \\
6432.3  &   0.29    & D &   Fe\,{\sc ii} (m40) 6432.7 \\
6442.7  &    0.49   &  D &   Fe\,{\sc ii} 6443.0 \\
6456.0 &   1.94    &   blend &   O\,{\sc i} (m9) 6456.0 \\
        &          &     &   Fe\,{\sc ii} (m74) 6456.4 \\
6465.5 &   0.06    &  S &  Uid \\
6483.4 &   0.41    &  S &  N\,{\sc i} (m21) 6482.7 \\
       &           &    &  N\,{\sc i} (m21) 6483.8 \\
       &            &    &  N\,{\sc i} (m21) 6484.9 \\
6492.2 &   1.04    &   blend &  Fe\,{\sc ii} 6491.3 \\
        &       &    &  Fe\,{\sc ii} 6493.1 \\
6506.1 &   0.35    &  D  &   Fe\,{\sc ii} 6506.3 \\
6516.2 &   1.25    &  M  &   Fe\,{\sc ii} (m40) 6516.1 \\
6547.5  &    0.05   & S  &   [N\,{\sc ii}] (m1F) 6548.1  \\
6562.9  &  194.30   & D  &   H$\alpha$ 6562.8 \\
6582.9  &  0.15   & S  &  [N\,{\sc ii}] (m1F) 6583.6  \\
6586.4 &   0.38    &  D & C\,{\sc i} (m22) 6587.8  \\
6598.5 &   0.22    &  S &  Uid \\
6613.5 &   0.15    & A  & Uid   \\
6626.5 &   0.12    & S  & Uid   \\
6658.3 &   0.09    &  S &  Uid \\
6666.2  &  0.21   & S  &  [Ni\,{\sc ii}] (m2F) 6666.8  \\
\hline
\end{tabular}
\end{table}

\begin{table}
\contcaption{ }
\begin{tabular}{cccc}
\hline
\hline
Wavelength (\AA)  &   W($\lambda$)  & Line Profile & Identification \\
\hline
6678.7  &   1.08    & S &  He\,{\sc i} (m46) 6678.2  \\
6716.0  &   0.08    & S &  [S\,{\sc ii}] (m2F) 6716.4  \\
6730.3 &    0.10   & S &   [S\,{\sc ii}] (m2F) 6730.8 \\
6776.3  &   0.08    & S &  Fe\,{\sc ii} (m210) 6777.3  \\
6814.3  &    0.12   & S &  C\,{\sc ii} 6812.2  \\
7001.8  &    0.22   & S & O\,{\sc i} (m21) 7001.9 \\
        &           &   & O\,{\sc i} (m21) 7002.2 \\
7041.9  &   0.14   &  S & Uid \\
7065.6  &    1.02   &  S & He\,{\sc i} (m10) 7065.2   \\
7109.1  &    0.10   & S &  Uid  \\
7154.6  &    0.47   &  D & [Fe\,{\sc ii}] (m14F) 7155.1 \\
7171.5  &    0.10   & S &   [Fe\,{\sc ii}] (m14F) 7171.9 \\
7308.1  &    0.15   &  S & Fe\,{\sc ii} (m73) 7308.0  \\
7320.1  &   0.16   &   S & [O\,{\sc ii}] (m2F) 7318.6 \\
7329.6  &   0.09   &  S  & [O\,{\sc ii}] (m2F) 7329.9 \\
7377.3  &    0.30   & S &   [Ni\,{\sc ii}] (m2F) 7377.9 \\
7387.2  &    0.33  & M &   [Fe\,{\sc ii}] (m14F) 7388.2 \\
7409.3  &    0.34   & M &  Fe\,{\sc ii} 7409.0 \\
7423.2  &    0.13   & S &  N\,{\sc i} (m3) 7423.6  \\
7442.0  &    0.20   & S &  N\,{\sc i} (m3) 7442.3  \\
7449.0  &    0.14   &  D & Fe\,{\sc ii} (m73) 7449.3  \\
7452.0  &    0.19   & S &   [Fe\,{\sc ii}] (m14F) 7452.5 \\
7461.9  &    0.57   &  M & Fe\,{\sc ii} (m73) 7462.4  \\
7467.7  &    0.25   & D &  N\,{\sc i} (m3) 7468.3  \\
7479.5  &    0.07   &  S & Fe\,{\sc ii} (m72) 7479.7  \\
7495.4 &    0.62   & D &   Uid  \\
7499.8  &    0.40   & D &   Uid  \\
7506.4  &    0.21   & D &   Uid  \\
7514.1 &    1.99   &  M & Fe\,{\sc ii} (m73) 7515.9  \\
7520.3 &    0.07   &  S & Fe\,{\sc ii} 7520.7  \\
7533.0  &    0.10   &  D & Fe\,{\sc ii} (m72) 7533.4  \\
7561.9  &    0.08   & A &   Uid  \\
7572.1  &    0.33   & M &   Fe\,{\sc ii} 7571.7  \\
7578.8  &    0.20  & D  & Uid \\
7587.4  &    0.07  & S & Uid \\
7711.2  &    2.32   & D &  Fe\,{\sc ii} (m73) 7711.7 \\
7731.6  &    0.43   & S &  Fe\,{\sc ii} 7731.7 \\
7755.2  &    0.20   & D &  Fe\,{\sc ii} 7755.6 \\
7764.1  &    0.07   & S &  [Fe\,{\sc ii}] (m30F) 7764.7 \\
7773.5 &    6.08   & blend &  O\,{\sc i} (m1) 7772.0 \\
         &       &  &  O\,{\sc i} (m1) 7774.2 \\
         &       &  &  O\,{\sc i} (m1) 7775.4 \\
7780.0 &    0.13   & S &  Fe\,{\sc ii} 7780.4 \\
7788.9 &    0.10   & D &  Fe\,{\sc ii} 7789.3 \\
7800.9 &    0.25   & D &  Fe\,{\sc ii} 7801.2 \\
7818.2  &    0.16   & D &  Fe\,{\sc ii} 7817.9 \\
7835.6  &    0.18   & M &  Fe\,{\sc ii} 7835.9 \\
7851.4  &    0.27   & S &  Fe\,{\sc ii} 7851.9 \\
7866.2  &    1.46   & D &  Fe\,{\sc ii} 7866.5 \\
7877.0  &    1.02   & D &  Mg\,{\sc ii} 7877.1 \\
7896.3  &    1.57   & D &  Mg\,{\sc ii} (m8) 7896.4 \\
7917.4  &    0.60   & D &  Fe\,{\sc ii} 7917.8 \\
7970.1  &    0.08   & S &  Fe\,{\sc ii} 7970.4 \\
7975.5  &    0.60   & D &  Fe\,{\sc ii} 7975.9 \\
7981.8  &    0.11   & S &  O\,{\sc i} 7982.0 \\
        &       &       &  O\,{\sc i} 7982.4 \\
8030.0  &    0.18   & S &  Fe\,{\sc ii} 8030.5 \\
8083.4  &    0.29   & D &  Fe\,{\sc ii} 8083.9 \\
8090.4  &    0.06   & D &  S\,{\sc ii} (m69) 8089.9 \\
        &           &   &  [Ti\,{\sc i}] (m18F) 8091.8 \\
8092.7  &    0.14   & D &  Fe\,{\sc ii} 8092.2 \\
8096.4  &    0.35  & M &  Uid \\
8101.2  &    0.19   & D &  Fe\,{\sc ii} 8101.5 \\
8106.4  &    0.42   & M &  [Cr\,{\sc ii}] (m20F) 8106.9 \\
\hline
\end{tabular}
\end{table}
                                                                                                                 
\begin{table}
\contcaption{ }
\begin{tabular}{cccc}
\hline
\hline
Wavelength (\AA)  &   W($\lambda$)  & Line Profile & Identification \\
\hline
8109.9  &    0.48  & D &  [Cr\,{\sc ii}] 8110.4 \\
8113.1  &    0.05  & S &  Uid \\
8114.9  &    0.11  & S &  Uid \\
8215.1  &    0.62   &  S &  N\,{\sc i} (m2) 8216.8 \\
8234.7  &    0.60   & S &  Mg\,{\sc ii} (m7) 8234.6 \\
8242.8  &    0.02   & S &  N\,{\sc i} (m2) 8242.3 \\
8249.9  &    0.38  & S &  P40 8250.0 \\
8267.7  &    0.17  & S &  P34 8267.9 \\
8292.0  &    0.31  & S &  P29 8292.3 \\
8298.6  &    0.83  & S &  P28 8298.8 \\
8323.5  &    1.13  & S &  P25 8323.4 \\
8333.6  &    1.49  & S &  P24 8334.0 \\
8345.2  &    2.05   &  S &  P23 8346.0 \\
8358.5 &     2.70   & S &   P22 8359.0 \\
8374.3  &    2.80   & S & P21 8374.5 \\
8392.2  &    3.34  &  S & P20 8392.4 \\
8413.1  &    3.62   & S & P19 8413.3    \\
8437.7  &    4.58   & S & P18 8438.0    \\
8445.9 &    35.62   & S &  O\,{\sc i} (m4) 8446.8 \\
8467.4  &    6.31   & S & P17 8467.3    \\
8489.8  &    1.72   & S &  Fe\,{\sc ii} 8490.1  \\
8501.7  &    9.96   & S & P16 8502.5    \\
8567.2  &    0.19   &  S & N\,{\sc i} (m8) 8567.7  \\
8598.2  &    10.77   & S & P14 8598.4    \\
8608.9  &    0.14   & S &  Fe\,{\sc ii} 8609.5  \\
8616.3  &    0.26   & S &  [Fe\,{\sc ii}] (m13F) 8617.0 \\
8628.9 &    1.08   &  D & N\,{\sc i} (m8) 8629.2  \\
8636.1  &    0.37   & S &  Fe\,{\sc ii} 8636.6  \\
8648.5  &    0.32   & S &  Uid \\
8655.4  &    0.19   & S & N\,{\sc i} (m8) 8655.9  \\
8664.4  &    12.00   & S & P13 8665.0    \\
8680.0 &    1.18   &  D & N\,{\sc i} (m1) 8680.2  \\
       &           &   & S\,{\sc i} (m6) 8680.5  \\
8682.9 &    0.88   &  S & N\,{\sc i} (m1) 8683.4  \\
8685.4 &    0.62   &  S & N\,{\sc i} (m1) 8686.1  \\
8694.6  &    0.27   & S & S\,{\sc i} (m6) 8694.7   \\
       &           &   & Fe\,{\sc ii} 8695.1   \\
8702.9 &    0.46   &  S & N\,{\sc i} (m1) 8703.2  \\
8711.2 &    0.45   &  S & N\,{\sc i} (m1) 8711.7  \\
8718.6 &    0.52   &  S & N\,{\sc i} (m1) 8718.8  \\
8721.9  &    0.42   & S &  Fe\,{\sc ii} 8722.4  \\
8728.1 &    0.25   &  S & [C\,{\sc i}] (m3F) 8727.4  \\
       &           &    & N\,{\sc i} (m1) 8728.9  \\
8750.2  &    11.50   & S & P12 8750.5    \\
8767.8  &    0.57   & S &  Uid \\
8774.4  &    0.09   & S &  Uid \\
8783.8  &    0.32   & S &  Uid \\
8805.6  &    0.57   & S &  Fe\,{\sc ii} 8805.1  \\
8813.6  &    0.38   & S &  Fe\,{\sc ii} 8813.4  \\
8819.3  &    0.35   & S &  Uid \\
8827.4  &    0.98   & S &  Fe\,{\sc ii} 8825.5 \\
         &       &  &  Fe\,{\sc ii} 8829.8 \\
8833.5  &    0.26   & S &  Fe\,{\sc ii} 8834.0 \\
8906.1  &    0.17   & S &  Uid \\
8911.4  &    0.62   & S &  Fe\,{\sc ii} 8912.5 \\
8915.9  &    0.29   & S &  Fe\,{\sc ii} 8916.3 \\
8926.2  &    2.93   & D &  Fe\,{\sc ii} 8926.7 \\
9014.6 &    11.80   &  S &  P10 9014.9 \\
9060.4  &    0.72   & S &  Fe\,{\sc ii} 9060.0 \\
      &           &    & N\,{\sc i} (m15) 9060.6  \\
      &           &    & C\,{\sc i} (m3) 9061.5  \\
      &           &    & C\,{\sc i} (m3) 9062.5  \\
9070.7  &    0.20   & S &  [S\,{\sc iii}] (m1F) 9069.4 \\
9075.8 &    2.11   & S &  Fe\,{\sc ii} 9075.5 \\
\hline
\end{tabular}
\end{table}

\begin{table}
\contcaption{ }
\begin{tabular}{cccc}
\hline
\hline
Wavelength (\AA)  &   W($\lambda$)  & Line Profile & Identification \\
\hline
9094.5  &    1.43   & S &  C\,{\sc i} (m3) 9094.9 \\
9111.5  &    0.68   & S &  C\,{\sc i} (m3) 9111.9 \\
9122.7  &    2.47  & D &  Fe\,{\sc ii} 9122.9 \\
9132.2  &    1.40  & S &  Fe\,{\sc ii} 9132.4 \\
9177.1  &    3.20  & D &  Fe\,{\sc ii} 9178.0 \\
9187.1  &    0.98  & S &  Fe\,{\sc ii} 9187.2 \\
9196.9  &    0.76  & S &  Fe\,{\sc ii} 9196.9 \\
9203.3  &    2.89  & S &  Fe\,{\sc ii} 9203.1 \\
\hline
\end{tabular}
\end{table}

\label{lastpage}

\end{document}